\documentclass{article} 
\usepackage{iclr2026_conference,times}

\usepackage{amsmath,amsfonts,bm}

\def\eqref#1{equation~\ref{#1}}

\def\1{\bm{1}}

\DeclareMathAlphabet{\mathsfit}{\encodingdefault}{\sfdefault}{m}{sl}
\SetMathAlphabet{\mathsfit}{bold}{\encodingdefault}{\sfdefault}{bx}{n}

\usepackage{hyperref}
\usepackage{url}
\usepackage{multirow} 
\usepackage{booktabs}
\usepackage{makecell}
\usepackage{tabularx}
\usepackage{siunitx,xcolor}
\usepackage{adjustbox}
\usepackage[table,dvipsnames]{xcolor}  
\usepackage{cleveref}
\usepackage{pifont}
\usepackage{listings}
\usepackage{caption}
\usepackage{graphicx}
\usepackage{epstopdf}
\usepackage{float}

\title{MR$^2$-Bench: Going Beyond Matching to Reasoning in Multimodal Retrieval}

\author{
  {\bf
    Junjie Zhou$^{1,2}$\thanks{Co-first authors.}
    \quad
    Ze Liu$^{3,2}$\footnotemark[1]
    \quad
    Lei Xiong$^{4,2}$
    \quad
    Jin-Ge Yao$^{2}$
    \quad
    Yueze Wang$^{2}$
    \vspace{4pt}
  } \\
  {\bf
    ~Shitao Xiao$^{2}$
    \enskip
    Fenfen Lin$^{2}$
    \enskip
    Miguel Hu Chen$^{2}$
    \enskip
    Zhicheng Dou$^{4}$
    \enskip
    Siqi Bao$^{5}$
    \vspace{4pt}
  } \\
  {\bf
    ~Defu Lian$^{3}$
    \enskip
    Yongping Xiong$^{1}$
    \enskip
    Zheng Liu$^{2}$\thanks{Corresponding author.}
    \vspace{4pt}
  } \\
  ~$^{1}$Beijing University of Posts and Telecommunications \\
  ~$^{2}$Beijing Academy of Artificial Intelligence \\
  ~$^{3}$University of Science and Technology of China \\
  ~$^{4}$Renmin University of China \quad
  $^{5}$Baidu Inc., China \\
  ~\texttt{\{junjiebupt, zhengliu1026\}@gmail.com}
  \hspace{12pt}
  \texttt{lz123@mail.ustc.edu.cn}
}

\begin{document}

\maketitle

\begin{abstract}
Multimodal retrieval is becoming a crucial component of modern AI applications, yet its evaluation lags behind the demands of more realistic and challenging scenarios. Existing benchmarks primarily probe surface-level semantic correspondence (e.g., object–text matching) while failing to assess the deeper reasoning required to capture complex relationships between visual and textual information. To address this gap, we introduce MR$^{2}$-Bench, a reasoning-intensive benchmark for multimodal retrieval. MR$^2$-Bench presents the following critical values: 1) all tasks are reasoning-driven, going beyond shallow matching to effectively assess models’ capacity for logical, spatial, and causal inference; 2) it features diverse multimodal data, such as natural images, diagrams, and visual puzzles, enabling comprehensive evaluation across content types; 
3) it supports complex queries and documents containing multiple images and covers diverse retrieval scenarios, more accurately reflecting real-world applications.
Our benchmark contains 1,309 curated queries, derived either from manual collection and annotation or from selective consolidation of public datasets. 
Despite achieving strong results on existing benchmarks, current state-of-the-art models still struggle on MR$^{2}$-Bench: for example, the leading Seed1.6-Embedding model attains a Recall@1 of 77.78 on MMEB, but only 9.91 on MR$^{2}$-Bench. This substantial performance gap highlights both the increased challenge posed by our benchmark and the pressing need for further advances in reasoning-intensive multimodal retrieval. The dataset and evaluation code will be made publicly available at \href{https://github.com/VectorSpaceLab/MR2-Bench}{https://github.com/VectorSpaceLab/MR2-Bench}.
\end{abstract}

\section{Introduction}
\label{intro}

Multimodal retrieval is a crucial capability in contemporary AI applications, supporting tasks such as image search~\citep{flickr-young2014image, magiclens}, retrieval-augmented generation (RAG)~\citep{murag-DBLP:conf/emnlp/ChenHCVC22,yu2024visrag}, and multimodal agentic systems~\citep{geng2025webwatcher, wu2025mmsearch-r1}. The field has evolved from traditional cross-modal matching (e.g., text-to-image retrieval~\citep{mscoco-chen2015microsoft}) to more advanced multimodal retrieval that accommodates compositional queries over interleaved image-text content (e.g., composed image retrieval~\citep{circo} and multimodal knowledge retrieval~\citep{chang2022webqa,remuq-DBLP:conf/acl/0003FGYB23}). Consequently, modern multimodal retrievers~\citep{vista,gme2025,vlm2vec-v2} can process queries expressed in text, images, or combinations thereof, efficiently extracting relevant information from diverse data sources and bridging the gap between complex datasets and real-world user needs.

Despite these advances, current evaluation methods remain misaligned with practical requirements. First, existing benchmarks primarily assess surface-level semantic correspondence, offering limited coverage of knowledge reasoning, spatial perception, and vision-centric challenges critical for diverse agentic applications. Second, these benchmarks predominantly feature natural images, with insufficient representation of visual puzzles, diagrams, and mathematical figures common in technical and educational contexts. Third, real-world documents often exhibit free-form, interleaved image-text layouts with multiple images positioned arbitrarily within the text. However, current benchmarks frequently limit each example to a single image~\citep{chang2022webqa, circo, OVEN-DBLP:conf/iccv/HuLCKJLTC23, jiangvlm2vec2025}, failing to reflect the complex document structures prevalent in practice. These limitations hinder rigorous evaluation of multimodal retrieval systems in reasoning-intensive, real-world scenarios.

In this paper, we introduce \textbf{MR$^{2}$-Bench} (\underline{M}ultimodal \underline{R}easoning-intensive \underline{R}etrieval \underline{Bench}mark). We summarize the key features of MR$^{2}$-Bench compared to existing benchmarks in \Cref{tab:benchmark-comparison}. In summary, MR$^{2}$-Bench presents the following critical advantages:

\vspace{-0.3cm}
\begin{itemize}
\item \textbf{It is the first benchmark for multimodal reasoning-intensive retrieval.} MR$^{2}$-Bench is pioneering in its requirement for reasoning to capture relevance rather than relying on shallow semantic matching, thereby filling a significant gap in current multimodal retrieval benchmarks. While existing text-only reasoning-intensive retrieval benchmarks \citep{rar-b2024,bright2025} have been developed, MR$^{2}$-Bench emphasizes multimodal capabilities with a variety of visually related reasoning-intensive retrieval tasks.

\item \textbf{It introduces a broad range of multimodal data domains.} Beyond typical natural images, MR$^{2}$-Bench incorporates diverse image types such as mathematical visual proofs, visual puzzles, and economic charts, etc. These images have widespread applications and inherently require visual reasoning capabilities. However, previous multimodal retrieval tasks have largely overlooked these data types.

\item \textbf{It offers diverse evaluation scenarios.} MR$^{2}$-Bench encompasses three meta-tasks: multimodal knowledge retrieval, visual illustration search, and visual relation reasoning, totaling 12 sub-tasks. These tasks provide a wide array of retrieval scenarios, including text-to-image, image-to-image, and mixed image-text queries, among others. Moreover, unlike previous multimodal benchmarks where queries or documents typically contain at most a single image~\citep{mbeir-wei2023uniir,jiangvlm2vec2025}, both queries and documents in MR$^{2}$-Bench may include multiple images, more accurately reflecting real-world scenarios.
\end{itemize}
\vspace{-0.2cm}

\definecolor{CheckGreen}{HTML}{2E7D32}
\definecolor{CrossRed}{HTML}{C62828}
\definecolor{PartAmber}{HTML}{F39C12}
\newcommand{\chkmark}{\textcolor{CheckGreen}{\ding{51}}}
\newcommand{\crossmark}{\textcolor{CrossRed}{\ding{55}}}

\begin{table}[t!]
\centering
{
\renewcommand{\arraystretch}{1.05}
\setlength{\extrarowheight}{0.18em}
\begin{adjustbox}{max width=\textwidth}
\begin{tabular}{lccccccc}
\toprule
\textbf{Benchmarks} & 
\textbf{\#Queries} & 
\textbf{\#Tasks} & 
\makecell{\textbf{Multi-}\\\textbf{Modality}} & 
\makecell{\textbf{Reasoning-}\\\textbf{Intensive}} & 
\makecell{\textbf{Vision-Centric}\\\textbf{Reasoning}} & 
\makecell{\textbf{Multi-}\\\textbf{Domain}} & 
\makecell{\textbf{Free-}\\\textbf{Form}} \\
\midrule
MS MARCO~\citep{bajaj2016ms}  &5,193 &1	&\crossmark & \crossmark & \crossmark & \crossmark & \crossmark \\
BEIR~\citep{muennighoff2022mteb}     & 54,262 & 18& \crossmark & \crossmark & \crossmark & \chkmark & \crossmark \\
RAR-b~\citep{rar-b2024}              & 45,745& 17 & \crossmark & \chkmark & \crossmark & \chkmark & \crossmark \\
BRIGHT~\citep{bright2025}            & 1,384 & 12 & \crossmark & \chkmark & \crossmark & \chkmark & \crossmark \\
\midrule
CIRR~\citep{cirr-liu2021image}       & 4,148 & 1 & \chkmark & \crossmark & \crossmark & \crossmark & \crossmark \\
WebQA~\citep{chang2022webqa}         & 7,540 & 1& \chkmark & \crossmark & \crossmark & \crossmark & \crossmark \\
M-BEIR~\citep{mbeir-wei2023uniir}    & 190,000 & 10& \chkmark & \crossmark & \crossmark & \chkmark & \crossmark \\
ViDoRe~\citep{colpali}               & 3,810 & 2 & \chkmark & \crossmark & \crossmark & \crossmark & \crossmark \\
MMEB~\citep{jiangvlm2vec2025}        & 36,000& 36& \chkmark & \crossmark & \crossmark & \chkmark & \crossmark \\
\midrule
\textbf{MR$^{2}$-Bench (Ours)}       & 1,309 & 12 & \chkmark & \chkmark & \chkmark & \chkmark & \chkmark \\
\bottomrule
\end{tabular}
\end{adjustbox}
}
\caption{Comparison of MR$^{2}$-Bench with existing benchmarks. Columns report the number of test queries (\textbf{\#Queries}); the number of tasks (\textbf{\#Tasks}); inclusion of image–text data (\textbf{Multi-Modality}); whether the benchmark is explicitly reasoning-focused (\textbf{Reasoning-Intensive}); whether it contains tasks solvable purely from images without textual cues (\textbf{Vision-Centric Reasoning}); domain coverage (\textbf{Multi-Domain}); and support for arbitrary text–image organization—interleaved ordering and multi-image on the query and document sides (\textbf{Free-Form}). The first block represents textual retrieval benchmarks, and the second block represents multimodal retrieval benchmarks.}
\vspace{-0.5cm}
\label{tab:benchmark-comparison}
\end{table}

We conduct comprehensive evaluation experiments on existing methods and derive the following key conclusions. Firstly, \textit{multimodal reasoning-intensive retrieval remains challenging for current retrievers}. Despite Seed1.6-Embedding~\citep{seed1-6-embedding} achieves the best performance on MR$^{2}$-Bench, it only reaches 30.68 nDCG@10. In contrast, it attains 77.78 Recall@1 on the MMEB dataset~\citep{jiangvlm2vec2025}, while its MR$^{2}$-Bench Recall@1 is just 9.91. Consistent failures are observed across all methods, particularly in mathematical visual proofs and visual relation reasoning.
Secondly, \textit{the capability of visual understanding plays an important role in solving our benchmark}. On the one hand, augmenting text-only retrievers with image captions yields substantial gains compared to ignoring images. On the other hand, despite current multimodal retrievers not being optimized for reasoning-intensive retrieval, the two strongest methods in our evaluation are native multimodal retrievers.
Finally, \textit{reasoning capacity holds significant potential for enhancing performance on MR²-Bench}. We implement reasoning-enhanced strategies including query rewriting and reranking, which have demonstrated substantial improvements on MR²-Bench.
These insights highlight the challenges and opportunities in multimodal retrieval. By exposing current strengths and weaknesses, we anticipate that MR$^2$-Bench will guide the development of more capable multimodal retrievers.

\section{Related Work}
\label{related-work}
\textbf{Reasoning-intensive Retrieval.} Information retrieval (IR) has advanced from lexical matching~\citep{bm25-robertson2009probabilistic} to capturing deep semantic relevance~\citep{dpr,bge,qwen3-embedding}. Recently, the rise of applications like retrieval-augmented generation and agentic systems~\citep{searcho1,searchr1,qian2025scent} has spurred the need for a more advanced capability: reasoning-intensive retrieval. This paradigm challenges IR systems to address complex information needs where relevance cannot be determined by direct semantic overlap, but must be inferred through deep reasoning. Although there has been significant progress in text-only domains with pioneering benchmarks such as BRIGHT~\citep{bright2025} and the development of specialized retrievers~\citep{shao2025reasonir,diver}, its application to multimodal scenarios remains largely unexplored. Our work addresses this gap for the first time. Beyond knowledge-oriented tasks, we introduce novel, vision-centric challenges, including visual illustration search and visual relational reasoning, requiring models to perform complex inference over integrated visual and textual data.

\textbf{Multimodal Retrieval.} As real-world information is increasingly presented in multimodal formats, multimodal retrieval has become essential for effectively searching corpora that integrate text and visual data. Initially, the focus was on cross-modal retrieval, such as text-to-image searches~\citep{mscoco-chen2015microsoft}. The field has since evolved to tackle more complex tasks, including image searches guided by textual instructions~\citep{fashioniq-wu2021fashion,magiclens}, multimodal document retrieval~\citep{chang2022webqa}, and knowledge retrieval using multimodal queries~\citep{remuq-DBLP:conf/acl/0003FGYB23}. With the advent of powerful pre-trained vision-language models (VLMs), researchers have been able to develop unified embedding models that effectively handle queries and documents in various formats~\citep{nv-mm-embed2024,megapairs}. Despite these advances, existing benchmarks and methods have largely concentrated on shallow semantic alignment or instance-level matching, neglecting the complex reasoning required to address many real-world information needs~\citep{mbeir-wei2023uniir,jiangvlm2vec2025}. Moreover, these benchmarks often emphasize natural images, overlooking visually complex and abstract domains that demand visual-centric reasoning abilities, such as visual puzzles, mathematical diagrams, and multi-image relational scenarios. Consequently, there is a pressing need for a benchmark designed to evaluate deeper reasoning capabilities in multimodal retrieval.

\section{MR$^2$-Bench: Multimodal Reasoning-intensive Retrieval Benchmark}
\label{dataset}

We propose MR$^2$-Bench, the first multimodal reasoning-intensive retrieval benchmark. A brief overview of MR$^2$-Bench's statistics is presented in Table~\ref{tab:stats}, and visual examples for each task type are shown in~\Cref{fig:dataset-example}. MR$^2$-Bench comprises 3 meta-tasks and 12 sub-tasks, encompassing a total of 1,309 queries. Detailed modalities of queries and documents, along with the instructions for each sub-task, are provided in Appendix~\ref{appendix:detail-overview}.

\begin{table}[htbp]
\centering
\begingroup
\setlength{\tabcolsep}{2.5pt}  
\scriptsize
\begin{adjustbox}{max width=\textwidth}
\begin{tabular}{@{}c|cccccc|ccc|ccc@{}}
\toprule
\textbf{Meta-task} & \multicolumn{6}{c|}{\textbf{Multimodal Knowledge Retrieval}} &
\multicolumn{3}{c|}{\textbf{Visual Illustration Search}} &
\multicolumn{3}{c}{\textbf{Visual Relation Reasoning}} \\ \midrule
\textbf{Sub-task} & \textbf{Biology} & \textbf{Cooking} & \textbf{Gardening} & \textbf{Physics} & \textbf{Chemistry} & \textbf{EarthScience} &
\textbf{Economics} & \textbf{Mathematics} & \textbf{Nature} & \textbf{Spatial} & \textbf{VisualPuzzle} & \textbf{Analogy} \\ \midrule
\textbf{\#Queries}  & 79 & 76 & 129 & 76 & 124 & 99 & 84 & 86 & 100 & 149 & 160 & 147 \\
\addlinespace[1pt]
\textbf{\#Corpus} & 4,455 & 2,786 & 5,636 & 6,656 & 4,317 & 3,014 & 7,572 & 944 & 2,017 & 1,000 & 5,375 & 3,970 \\
\bottomrule
\end{tabular}
\end{adjustbox}
\endgroup
\caption{Data statistics of queries and corpus for each sub-task in MR\(^2\)-Bench}
\vspace{-0.5cm}
\label{tab:stats}
\end{table}

\subsection{Multimodal Knowledge Retrieval}

Traditional knowledge retrieval has focused primarily on text-only queries and corpora~\citep{chen-etal-2017-reading, kwiatkowski-etal-2019-natural}. However, images play a crucial role in realistic knowledge retrieval scenarios. For instance, when users wish to explore an intriguing scientific phenomenon in their daily lives, capturing an image for querying is often more intuitive and detailed than using text alone. Similarly, knowledge bases frequently integrate text and images, with images providing essential explanatory and knowledge representation functions. Although some benchmarks have been developed for multimodal knowledge search~\citep{chang2022webqa, remuq-DBLP:conf/acl/0003FGYB23, OVEN-DBLP:conf/iccv/HuLCKJLTC23, infoseek-bench-2023}, they are predominantly based on annotations from sources like Wikidata, with questions that are often straightforward (e.g., \textit{What is this mountain called?}\footnote{Query example curated from the OVEN benchmark~\citep{OVEN-DBLP:conf/iccv/HuLCKJLTC23}}). These tasks typically rely on keyword matching, image instance matching, or simple shallow semantic alignment. However, real-world user queries can be highly complex, requiring intensive reasoning to identify relevant documents.

BRIGHT~\citep{bright2025} introduced the first benchmark for evaluating reasoning-intensive knowledge retrieval by constructing retrieval pairs between real user queries from Stack Exchange\footnote{\href{https://stackexchange.com}{https://stackexchange.com}} and relevant documents. The relevant documents are identified from external links referenced in high-scoring answers, establishing retrieval relationships that require reasoning over critical concepts or theories to bridge the query and the document. As a result, retrieval models evaluated on this benchmark must possess capabilities that go beyond simple lexical or semantic matching. However, BRIGHT is a text-only benchmark, leaving a gap in multimodal queries and documents.

\begin{figure}[htbp]
    \centering
    \includegraphics[width=\linewidth]{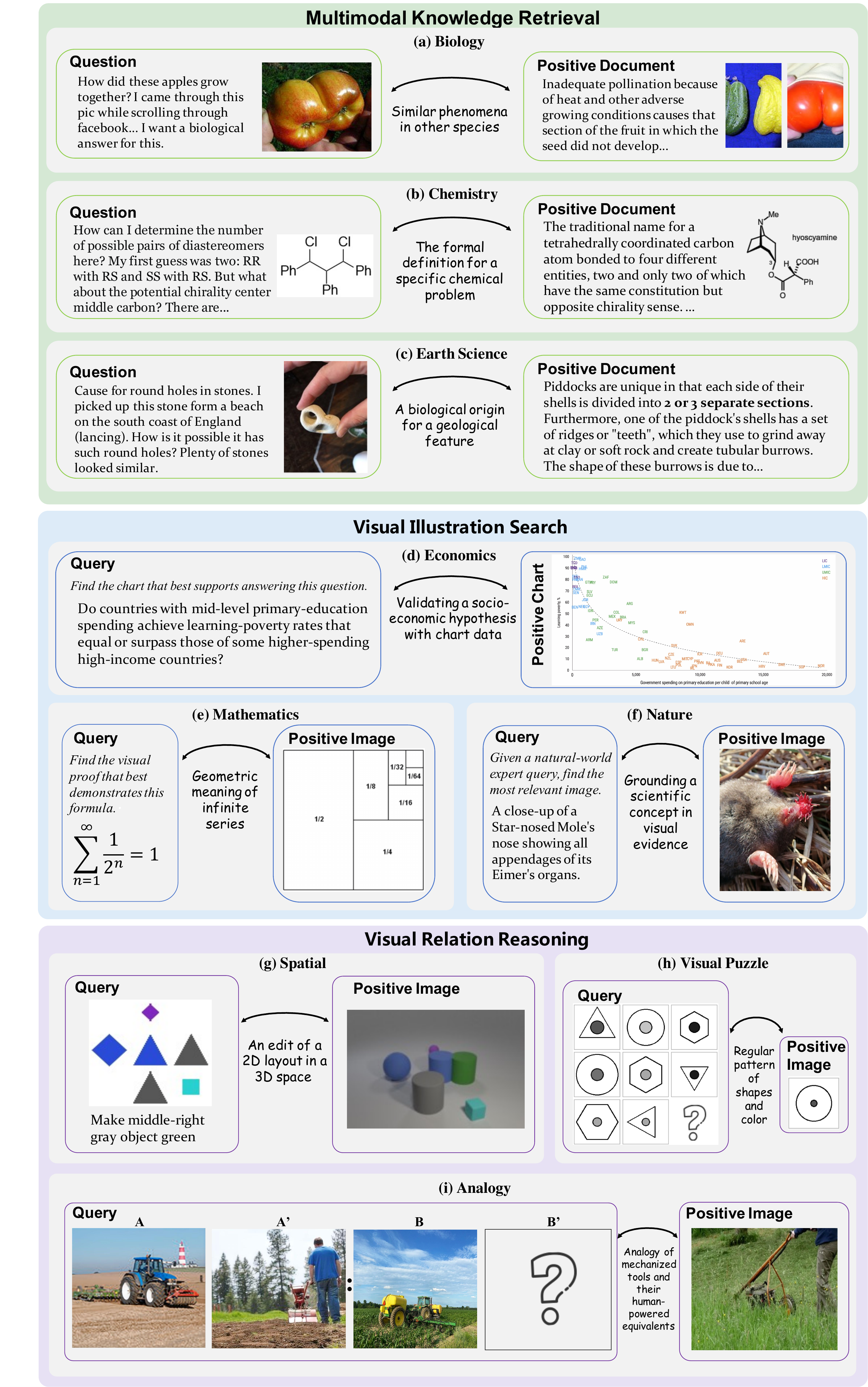}
    \caption{Visualized Examples of MR\(^2\)-Bench: Sub-task illustrations from three meta-tasks, with 3 out of 6 shown for the multimodal knowledge retrieval task.}    
    \label{fig:dataset-example}
\end{figure}

Inspired by BRIGHT's task construction approach, we have developed a set of reasoning-intensive multimodal knowledge retrieval tasks in our MR\(^2\)-Bench. In contrast to BRIGHT, our approach rigorously ensures that images are essential components of the questions, rendering these inquiries invalid without the accompanying visual data. We also retain images from relevant documents if they are crucial for conveying knowledge. The annotation process is detailed in the Appendix~\ref{appendix:data-construction}. Our benchmark covers six domains: \textbf{Biology}, \textbf{Cooking}, \textbf{Gardening}, \textbf{Physics}, \textbf{Chemistry}, and \textbf{Earth Science}. Examples of these tasks are illustrated in~\Cref{fig:dataset-example}(a)-(c). For instance, in~\Cref{fig:dataset-example}(a), the positive document does not mention \textit{apple} or \textit{grow together}. The key to connecting the document and the question lies in the accompanying image, which demonstrates a similar biological phenomenon in other species.

\subsection{Visual Illustration Search}

Text-to-image retrieval (e.g., Flickr30K~\citep{flickr-young2014image}, MSCOCO~\citep{mscoco-chen2015microsoft}) is a canonical multimodal retrieval task, where the system need to retrieve the image that best matches a textual query. Classic benchmarks are largely limited to direct and surface-level semantic alignment, such as identifying a specific animal or a person performing a certain sport.
However, real-world use cases often require domain knowledge and multi-step reasoning to retrieve the target image (e.g., professional charts and scientific illustrations). To address this gap, we introduce the \textbf{Visual Illustration Search} (VIS) task. In this task, the model is required to retrieve an image that functions as a visual illustration, intuitively explaining or solving a problem posed in a challenging, domain-specific textual query. Comprising three sub-tasks: \textbf{Economics}, \textbf{Mathematics}, and \textbf{Nature}, VIS evaluates a model’s ability to perform cross-modal reasoning and knowledge-grounded understanding in complex multimodal scenarios.

\textbf{Economics}. Charts serve as intuitive illustrations across various disciplines. 
However, existing chart-related tasks (e.g., ChartVQA~\citep{masry-etal-2022-chartqa}, ViDoRe~\citep{colpali}) primarily test surface-level abilities solvable with basic OCR and arithmetic. 
To assess a model's ability to capture the deeper semantics and domain knowledge embedded in chart, we manually collected reports from the World Bank\footnote{\href{https://data.worldbank.org}{https://data.worldbank.org}}, extracted charts related to economics, and asked human experts to create questions grounded in these charts.
The core annotation principle is that each question must demand sufficient reasoning to identify the positive chart.
For instance, as shown in Figure~\ref{fig:dataset-example}(d), the positive chart does not explicitly state the conclusion;
only by comparing the relative positions of different countries in the chart and associating \textit{spending quantiles} with \textit{learning poverty rates} can one validate the hypothesis posed in the question. 
Following these principles, we constructed a reasoning-oriented retrieval subset centered on economic charts, comprising 84 high-quality questions.

\textbf{Mathematics}. Images can effectively reinforce human  comprehension of abstract knowledge. This holds especially in mathematics, where \textit{visual proofs} are conical examples that use geometric relations to demonstrate abstract theorems intuitively. 
As shown in Figure~\ref{fig:dataset-example}(e), the recursive partition of the unit square gives a clear proof of the infinite series $\sum_{n=1}^{\infty} \frac{1}{2^n}=1$.
Although structurally simple, such proofs embody rigorous logic and require strong reasoning to connect visual patterns with abstract mathematical principles, providing an effective evaluation of model’s reasoning ability.
However, visual proofs are largely absent from existing multimodal retrieval benchmarks.
Therefore, we curate 86 mathematical formulas from \emph{Proofs Without Words}~\citep{nelsen2015proofs} and 
Wikimedia Commons\footnote{\href{https://commons.wikimedia.org/wiki/Category\%3AProof\_without\_words}{https://commons.wikimedia.org/wiki/Category\%3AProof\_without\_words}}, using each formula as a query and its corresponding visual proof as the positive image.

\textbf{Nature}. Natural-world images are more than depictions; they are visual reference for species identification, ecosystem monitoring, and science education~\citep{van2015building,van2018inaturalist}, which require images that capture specific traits or morphology, rather than the generic picture of the organism. For example, as shown in Figure~\ref{fig:dataset-example}(f), the query seeks for \textit{a close-up of star-nosed mole’s distinctive organs}, which demands both expert biological knowledge and fine-grained visual recognition.  
Satisfying such knowledge-intensive visual requests is a challenging yet essential capability for models. To evaluate this, we carefully selected 100 queries from the publicly available INQUIRE-Rerank dataset~\citep{vendrow2024inquire} to construct the expert-level natural-world image retrieval task.

\subsection{Visual Relation Reasoning}

In prevailing multimodal retrieval benchmarks, textual queries are the primary driver of user intent.
However, this paradigm often overlooks the rich, self-contained semantics inherent in purely visual structures and relationships that are independent of natural language. To address this gap, we introduce \textbf{Visual Relation Reasoning}, a suite of tasks for assessing high-level vision-centric reasoning through three distinct sub-tasks: \textbf{Spatial}, \textbf{Visual Puzzle}, and \textbf{Analogy}.

\textbf{Spatial}. The capacity for spatial perception, transformation, and reasoning is essential for models.
To evaluate these capabilities, we incorporate tasks from the CSS dataset~\citep{vo2019composing}, a controlled synthetic dataset where each sample consists of a reference image, a textual modification instruction, and a corresponding target image, with scenes rendered as both 2D layouts and photorealistic 3D images. As illustrated in Figure~\ref{fig:dataset-example}(g), the query requires jointly parsing descriptions that combine relative position and attributes (i.e., \textit{middle-right gray object}) and projecting the 2D layout into the corresponding 3D scene, yielding a comprehensive test of spatial ability. From CSS, we curated 149 queries to constitute the spatial-reasoning subtask of MR$^{2}$-Bench.

\textbf{Visual Puzzle}. Inspired by Raven's Progressive Matrices\footnote{\href{https://en.wikipedia.org/wiki/Raven\%27s\_Progressive\_Matrices}{https://en.wikipedia.org/wiki/Raven\%27s\_Progressive\_Matrices}}, this task is designed to evaluate pattern recognition and structural reasoning.  As shown in Figure~\ref{fig:dataset-example}(h), for a given 3×3 matrix with the final cell missing, the model need to retrieve the positive image that logically completes the matrix's underlying pattern. This task is distinguished by its near-complete absence of linguistic signals, which compels the model to directly infer abstract patterns 
to perform higher-order reasoning from vision alone. We reorganized the RAVEN dataset~\citep{zhang2019raven}: for each rule-governed visual attribute, we selected a set of queries, pooled the corresponding candidate images and removed duplicates to build the corpus. In total, we curated 160 queries for this task.

\textbf{Analogy}. Derived from the VASR dataset~\citep{bitton2023vasr}, this task tests a model's capability for visual analogical reasoning. 
As shown in~\Cref{fig:dataset-example}(i), the query comprises three images $(A, A', B)$, where the pair $(A, A')$ exemplifies a visual semantic transformation (e.g., \textit{replacing a machine with human labor in a comparable scene}) that is expected to hold between $B$ and $B'$.
The model must infer the transformation from $A$ to $A'$, apply it to $B$, and retrieve the image $B'$ that completes the analogy. 
It requires the model abstracts an implicit transformation rule from one image pair and generalizes it to another, which effectively tests its capacity for high-order visual reasoning. We instantiate this task by converting VASR analogy triplets into a retrieval setting and curated 147 challenging queries.

\section{Experiments}
\subsection{Settings}

We evaluated 11 popular embedding models using our MR\(^2\)-Bench, categorizing them into two main types: text-only embedding models and multimodal embedding models. We employed nDCG@10 as the primary metric, with additional metric results provided in Appendix~\ref{appendix:detailed-evaluation}. 

For \textit{text embedding models}, we assessed two categories: traditional models such as BGE-M3~\citep{bge-m3} and Qwen3-Embedding~\citep{qwen3-embedding}, and models optimized for reasoning-intensive retrieval, including ReasonIR~\citep{shao2025reasonir}, BGE-Reasoner-Embed\footnote{https://huggingface.co/BAAI/bge-reasoner-embed-qwen3-8b-0923}, and Diver-Embed~\citep{diver}. We adopted two evaluation approaches for text embedding models: (1) Using only text information from queries and documents, which is limited for tasks where queries or candidates are purely image-based; (2) Replacing images with textual descriptions (captions). 
For \textit{multimodal embedding models}, we evaluated CLIP~\citep{clip-radford2021learning}, VISTA~\citep{vista}, BGE-VL~\citep{megapairs}, MM-Embed~\citep{nv-mm-embed2024}, GME~\citep{gme2025}, VLM2VecV2~\citep{vlm2vec-v2}, and Seed1.6-Embedding~\citep{seed1-6-embedding}. Detailed information on the models and evaluation procedures can be found in Appendix~\ref{appendix:baselines}.

\subsection{Main Results}

\definecolor{caprow}{HTML}{F2F2F2}
\begin{table}[t]
\begin{adjustbox}{max width=\textwidth}
{
\renewcommand{\arraystretch}{1.10}   
\setlength{\extrarowheight}{0.2em}
\begin{tabular}{@{}cccccccccccccc@{}}
\toprule
\multicolumn{1}{c|}{\multirow{2}{*}{\textbf{Methods}}} & \multicolumn{6}{c|}{\textbf{Multimodal Knowledge Retrieval}} & \multicolumn{3}{c|}{\textbf{Visual Illustration}} & \multicolumn{3}{c|}{\textbf{Visual Relation}} & \multicolumn{1}{l}{\multirow{2}{*}{\textbf{Avg.}}} \\ \cmidrule(lr){2-7} \cmidrule(lr){8-10} \cmidrule(lr){11-13}
\multicolumn{1}{c|}{} & \textbf{Bio.} & \textbf{Cook.} & \textbf{Gar.} & \textbf{Phy.} & \textbf{Chem.} & \multicolumn{1}{c|}{\textbf{Earth.}} & \textbf{Econ.} & \textbf{Math.} & \multicolumn{1}{c|}{\textbf{Nat.}} & \textbf{Spa.} & \textbf{Puzz.} & \multicolumn{1}{c|}{\textbf{Ana.}} & \multicolumn{1}{l}{} \\ \midrule
\multicolumn{14}{c}{\textcolor{gray}{\textbf{\textit{Text Embedding Models}}}} \\ \midrule
\multicolumn{1}{l|}{BGE-M3} & 18.79 & 12.97 & 12.04 & 14.52 & 6.05 & \multicolumn{1}{c|}{16.35} & - & - & \multicolumn{1}{c|}{-} & - & - & \multicolumn{1}{c|}{-} & - \\
\rowcolor{caprow}
\multicolumn{1}{r|}{\textit{+ Captions}}& {34.19} & {24.28} & {17.88} & {21.24} & {9.67} & \multicolumn{1}{c|}{{25.19}} & {45.46} & {9.97} & \multicolumn{1}{c|}{{23.66}} & {9.48} & {0.00} & \multicolumn{1}{c|}{{3.46}} & {18.71} \\
\addlinespace[2pt]
\multicolumn{1}{l|}{Qwen3} & 23.77  & 20.44 & 12.61 & 17.13 & 8.61 & \multicolumn{1}{c|}{19.79} & - & - & \multicolumn{1}{c|}{-} & - & - & \multicolumn{1}{c|}{-} & - \\
\rowcolor{caprow}
\multicolumn{1}{r|}{\textit{+ Captions}} & {29.97} & {29.29} & {18.32} & {21.46} & {9.52} & \multicolumn{1}{c|}{{23.19}} & {49.44} & {21.14} & \multicolumn{1}{c|}{{26.30}} & {9.11} & {0.00} & \multicolumn{1}{c|}{{4.30}} & {20.17} \\
\addlinespace[2pt]
\multicolumn{1}{l|}{Diver-Emb.} & 27.32 & 16.94 & 15.17 & 18.05 & 10.06 & \multicolumn{1}{c|}{22.57} & - & - & \multicolumn{1}{c|}{-} & - & - & \multicolumn{1}{c|}{-} & - \\
\rowcolor{caprow}
\multicolumn{1}{r|}{\textit{+ Captions}} & 38.46 & 30.87 & 22.84 & 23.62 & 14.46 & \multicolumn{1}{c|}{31.40} & 54.67 & 25.91 & \multicolumn{1}{c|}{24.88} & 8.52 & 0.00 & \multicolumn{1}{c|}{7.47} & 23.59 \\
\addlinespace[2pt]
\multicolumn{1}{l|}{BGE-Rea.} & 29.01 & 15.37 & 16.31 & 21.00 & 10.62 & \multicolumn{1}{c|}{26.20} & - & - & \multicolumn{1}{c|}{-} & - & - & \multicolumn{1}{c|}{-} & - \\
\rowcolor{caprow}
\multicolumn{1}{r|}{\textit{+ Captions}} & 42.60 & 34.40 & \underline{24.94} & 25.61 & 14.31 & \multicolumn{1}{c|}{34.57} & 54.31 & 17.16 & \multicolumn{1}{c|}{29.86} & 5.52 & 0.00 & \multicolumn{1}{c|}{5.88} & 25.35 \\
\addlinespace[2pt]
\multicolumn{1}{l|}{ReasonIR} & 29.85 & 19.72 & 16.22 & 21.56 & 9.83 & \multicolumn{1}{c|}{23.56} & - & - & \multicolumn{1}{c|}{-} & - & - & \multicolumn{1}{c|}{-} & - \\
\rowcolor{caprow}
\multicolumn{1}{r|}{\textit{+ Captions}} & \underline{44.75} & \underline{41.91} & 18.79 & 27.33 & \underline{17.45} & \multicolumn{1}{c|}{\underline{41.22}} & \textbf{64.04} & \textbf{34.49} & \multicolumn{1}{c|}{30.70} & 11.65 & 0.00 & \multicolumn{1}{c|}{\underline{10.89}} & 25.72 \\
 \midrule
\multicolumn{14}{c}{\textcolor{gray}{\textbf{\textit{Multimodal Embedding Models}}}} \\ \midrule
\multicolumn{1}{l|}{CLIP} & 32.85 & 30.57 & 14.06 & 14.86 & 3.50 & \multicolumn{1}{c|}{33.23} & 12.97 & 5.64 & \multicolumn{1}{c|}{\underline{49.34}} & \underline{20.89} & 0.19 & \multicolumn{1}{c|}{5.09} & 18.59 \\
\multicolumn{1}{l|}{BGE-VL} & 29.41 & 18.36 & 10.50 & 19.51 & 7.12 & \multicolumn{1}{c|}{19.73} & 50.80 & 14.31 & \multicolumn{1}{c|}{47.97} & 6.46 & 0.00 & \multicolumn{1}{c|}{0.75} & 19.53 \\
\multicolumn{1}{l|}{GME} & 34.34 & 39.50 & 19.04 & 19.29 & 7.73 & \multicolumn{1}{c|}{28.59} & 36.95 & 7.19 & \multicolumn{1}{c|}{39.35} & 15.70 & 0.22 & \multicolumn{1}{c|}{\textbf{11.11}} & 21.59 \\
\multicolumn{1}{l|}{VLM2Vec} & 39.37 & 39.38 & 19.87 & 20.28 & 9.03 & \multicolumn{1}{c|}{35.71} & 51.44 & 14.16 & \multicolumn{1}{c|}{35.06} & 13.94 & \underline{0.62} & \multicolumn{1}{c|}{5.85} & 23.72 \\
\multicolumn{1}{l|}{MM-Emb.} & \textbf{49.68} & \textbf{52.19} & 23.67 & \textbf{30.36} & 17.44 & \multicolumn{1}{c|}{\textbf{47.51}} & 42.99 & 21.58 & \multicolumn{1}{c|}{48.41} & \textbf{22.79} & 0.21 & \multicolumn{1}{c|}{5.93} & \underline{30.23} \\
\multicolumn{1}{l|}{Seed-1.6} & 40.64 & 38.12 & \textbf{31.77} & \underline{27.91} & \textbf{17.80} & \multicolumn{1}{c|}{37.17} & \underline{56.13} & \underline{26.10} & \multicolumn{1}{c|}{\textbf{65.16}} & 17.29 & \textbf{0.93} & \multicolumn{1}{c|}{9.21} & \textbf{30.68} \\ 
\bottomrule
\end{tabular}
}
\end{adjustbox}
\caption{\textbf{The overall performance of embedding models on MR$^{2}$-Bench.} We report nDCG@10 for all sub-tasks. Avg. denotes the average score across 12 datasets. The best score on each dataset is shown in bold and the second best is underlined.}
\vspace{-0.5cm}
\label{tab:main-results}
\end{table}
We summarize the overall evaluation results for all investigated retrieval baselines in MR$^2$-Bench in \Cref{tab:main-results}. For each sub-task, we report nDCG@10, along with the macro-average (Avg.) across all tasks. From these results, we draw some primary conclusions:

\textbf{1) Current state-of-the-art models underperform on MR$^2$-Bench.} The leading Seed-1.6 Embedding model~\citep{seed1-6-embedding} achieves only 30.68 nDCG@10 on our benchmark. In contrast, it reports 77.78 overall Recall@1 on the popular MMEB leaderboard~\citep{jiangvlm2vec2025}, but its performance drops significantly to 9.91 Recall@1 on MR$^2$-Bench. Additionally, the SOTA reasoning-intensive text retriever, Diver-Retriever~\citep{diver}, achieves 33.90 nDCG@10 on BRIGHT~\citep{bright2025}, yet only reaches 23.59 nDCG@10 on MR$^2$-Bench when evaluated with auxiliary captions. These results highlight the increased challenges posed by our MR$^2$-Bench.

\textbf{2) Text retrievers augmented with image captions provide a strong and practical baseline on MR$^2$-Bench.} Since text retrievers cannot directly process images, we replace each image in queries and candidate documents with detailed natural-language descriptions. This augmentation leads to notable improvements. For instance, ReasonIR\textit{+Captions} surpasses popular open-source multimodal retrievers like VLM2Vec-V2~\citep{vlm2vec-v2}. On the Stack Exchange subset, adding captions consistently boosts performance across most tasks. These findings confirm that MR$^2$-Bench is fundamentally multimodal, with retrieval performance significantly enhanced by the visual information provided through captions.

\textbf{3) Reasoning-oriented text retrievers significantly outperform traditional matching-based retrievers.} Models optimized for reasoning-intensive retrieval, such as ReasonIR and Diver-Retriever, consistently achieve higher nDCG@10 scores on MR$^2$-Bench compared to matching-centric retrievers like BGE-M3 and Qwen3-Embedding. This advantage is evident across various meta-tasks and persists whether visual content is absent or represented as detailed captions. Collectively, these findings suggest that reasoning-oriented capabilities learned in text retrieval effectively transfer to multimodal retrieval tasks requiring complex reasoning.

\textbf{4) Multimodal retrievers show potential on MR$^2$-Bench.} Although not specifically designed for reasoning-intensive tasks, multimodal embedding models like MM-Embed and Seed1.6-Embedding lead performance on MR$^2$-Bench. These models notably outperform caption-augmented text retrievers, including those optimized for reasoning. 
This gap suggests a promising direction for future research in developing reasoning-intensive multimodal retrievers.

\textbf{5) Existing methods struggle with capturing complex visual relationships and abstract concepts.} Current models face challenges in effectively perceiving multi-image relationships (Analogy), spatial configurations (Spatial), and abstract graphics (Mathematics, Visual Puzzle). We hypothesize that these difficulties stem from the inherently visual-centric nature of these tasks, which existing embedding models struggle to comprehend fully. Nonetheless, these images are crucial for real-world applications, as their information is difficult to convey through language alone. This indicates substantial potential for future research to enhance multimodal embedding models.

\vspace{-0.2cm}
\subsection{More Analysis}
\vspace{-0.1cm}

\subsubsection{The Effectiveness of Query Rewriting}
\vspace{-0.1cm}

\textbf{6) Query rewriting enhances both text and multimodal baselines on MR\(^2\)-Bench.} This generation-augmented retrieval technique clarifies complex user intent and highlights latent constraints, thus facilitating reasoning-intensive retrieval. Although extensively studied in text-only contexts \citep{hyde2023, reinforced-ir}, its application to multimodal retrieval remains underexplored. We evaluated a simple, model-agnostic query rewriting pipeline on MR\(^2\)-Bench. For each query, GPT-5~\citep{GPT-5} generates step-by-step reasoning, which is then utilized by each retriever (details in Appendix~\ref{appendix:query-rewriting}). As shown in \Cref{tab:re-write}, both text and multimodal retrievers show notable average improvements. These results indicate that query rewriting is a practical method for enhancing multimodal reasoning-intensive retrieval tasks, consistently improving performance without the need for fine-tuning existing retrievers.

\definecolor{caprow}{HTML}{F2F2F2}
\begin{table}[htb]
\begin{adjustbox}{max width=\textwidth}
{
\renewcommand{\arraystretch}{1.10}   
\setlength{\extrarowheight}{0.2em}
\begin{tabular}{@{}cccccccccccccc@{}}
\toprule
\multicolumn{1}{c|}{\multirow{2}{*}{\textbf{Methods}}} & \multicolumn{6}{c|}{\textbf{Stack Exchange}} & \multicolumn{3}{c|}{\textbf{Visual Illustration}} & \multicolumn{3}{c|}{\textbf{Visual Relation}} & \multicolumn{1}{l}{\multirow{2}{*}{\textbf{Avg.}}} \\ \cmidrule(lr){2-7} \cmidrule(lr){8-10} \cmidrule(lr){11-13}
\multicolumn{1}{c|}{} & \textbf{Bio.} & \textbf{Cook.} & \textbf{Gar.} & \textbf{Phy.} & \textbf{Chem.} & \multicolumn{1}{c|}{\textbf{Earth.}} & \textbf{Econ.} & \textbf{Math.} & \multicolumn{1}{c|}{\textbf{Nat.}} & \textbf{Spa.} & \textbf{Puzz.} & \multicolumn{1}{c|}{\textbf{Ana.}} & \multicolumn{1}{l}{} \\ \midrule
\multicolumn{1}{l|}{BGE-M3} & {34.19} & {24.28} & {17.88} & {21.24} & {9.67} & \multicolumn{1}{c|}{{25.19}} & {45.46} & {9.97} & \multicolumn{1}{c|}{{23.66}} & {9.48} & {0.00} & \multicolumn{1}{c|}{{3.46}} & {18.71} \\
\rowcolor{caprow}
\multicolumn{1}{r|}{\textit{+ Rewrite}}& {40.41} & {32.94} & {25.66} & {23.12} & {11.98} & \multicolumn{1}{c|}{{33.63}} & {50.88} & {20.09} & \multicolumn{1}{c|}{{23.38}} & {7.13} & {0.00} & \multicolumn{1}{c|}{{7.91}} & {\textbf{23.09}} \\
\addlinespace[2pt]
\multicolumn{1}{l|}{Seed-1.6} & 40.64 & 38.12 & {31.77} & {27.91} & {17.80} & \multicolumn{1}{c|}{37.17} & {56.13} & {26.10} & \multicolumn{1}{c|}{{65.16}} & 17.29 & {0.93} & \multicolumn{1}{c|}{9.21} & {30.68} \\
\rowcolor{caprow}
\multicolumn{1}{r|}{\textit{+ Rewrite}} & 41.13 & 41.47 & 37.68 & 29.47 & 20.70 & \multicolumn{1}{c|}{42.02} & 50.08 & 30.37 & \multicolumn{1}{c|}{65.84} & 31.87 & 1.24 & \multicolumn{1}{c|}{14.62} & \textbf{33.87} \\
\addlinespace[2pt]
\bottomrule
\end{tabular}
}
\end{adjustbox}
\caption{Performance comparison of BGE-M3 and Seed-1.6 Embedding on MR$^2$-Bench before and after query rewriting, showing significant improvements across most tasks.} %
\vspace{-0.5cm}
\label{tab:re-write}
\end{table}

\subsubsection{The Effectiveness of Advanced Reranking}
A common approach to improve retrieval performance is to employ rerankers that jointly process both the query and its retrieved candidates. Existing studies have shown that incorporating an intermediate reasoning step before final scoring can lead to more accurate rankings~\citep{weller2025rank1testtimecomputereranking,zhuang2025rank-r1,liu2025reasonrank}.
We also investigate this by incorporating a reranking stage after the initial retrieval on MR$^{2}$-Bench. Specifically, we test a wide range of rerankers to rerank the top-$k=20$ candidates retrieved by three base retrievers: Qwen3-Embedding, GME, and Seed-1.6-Embedding. Their retrieved candidates are reanked by: 1) \textit{textual rerankers}: RankLLaMA-7B and RankLLaMA-14B~\citep{ma2024fine}; 2) \textit{reasoning-enhanced textual rerankers}: Rank1-7B~\citep{weller2025rank1testtimecomputereranking}, RankR1-14B~\citep{zhuang2025rank-r1}, ReasonRank-32B~\citep{liu2025reasonrank}, and BGE-Reasoner-Reranker-32B~\footnote{\href{https://github.com/FlagOpen/FlagEmbedding/tree/master/research/BGE_Reasoner}{https://github.com/FlagOpen/FlagEmbedding/tree/master/research/BGE\_Reasoner}}; 3) \textit{multimodal rerankers}: MonoQwen2-VL-v0.1~\citep{MonoQwen} and Jina-Reranker-m0~\citep{jina-reranker-m0}; and 4) \textit{reasoning-enhanced multimodal rerankers}: Gemma3-27B~\citep{gemma_2025}, Qwen2.5-VL-72B~\citep{Qwen2.5-VL}, GLM-4.5V~\citep{vteam2025glm45vglm41vthinkingversatilemultimodal}, and GPT-5~\citep{GPT-5}. Since there are no off-the-shelf multimodal rerankers that natively support reasoning, we prompt these MLLMs to first perform reasoning and then output a relevance score. Full implementation details are available in Appendix~\ref{appendix: rerank settings}. Average performance based on Seed-1.6-Embedding is shown in Figure~\ref{fig:seed_embedding_reranker}, and detailed results for all three base retrievers are provided in the Appendix~\ref{appendix: rerank reuslts}. 

\begin{figure}[htbp]
    \centering
    \includegraphics[width=\linewidth]{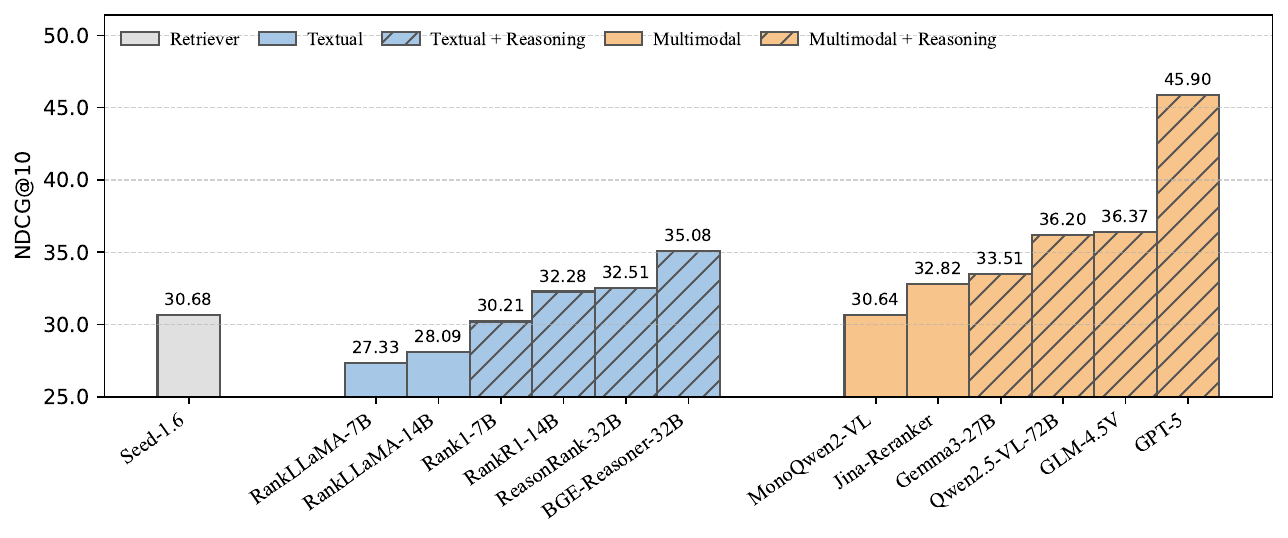}
    \vspace{-2em}
    \caption{Reranking performance on MR$^{2}$-Bench with Seed-1.6-Embedding as the base retriever.}
    \vspace{-0.3cm}
    \label{fig:seed_embedding_reranker}
\end{figure}

From the results presented in Figure~\ref{fig:seed_embedding_reranker}, we have following findings:

\textbf{7) Rerankers deliver substantial gains on MR$^{2}$-Bench.} Most rerankers significantly outperform the strong Seed-1.6-Embedding baseline, demonstrating the benefit of joint modeling of queries and candidates. Notably, GPT-5 achieves an nDCG@10 of 45.90, an absolute gain of 15.22 over the baseline, indicating the substantial headroom for improvement unlocked by reranking.

\textbf{8) An explicit reasoning step before scoring proves to be beneficial.} Across text-only rerankers, those incorporating reasoning consistently outperform their non-reasoning, size-matched counterparts (e.g., Rank1-7B vs. RankLLaMA-7B; RankR1-14B vs. RankLLaMA-14B). This is further substantiated by BGE-Reasoner-Reranker-32B: using only textual input, it achieves an nDCG@10 of 35.08, outperforming the strong base retriever by 4.2 points. Moreover, for multimodal rerankers, models prompted to reason and then rank outperform those trained non-reasoning rerankers. These results confirm that explicit reasoning drives the gains on MR$^{2}$-Bench.

\textbf{9) Multimodal information plays a significant role in enhancing performance.} Despite being built on the lightweight Qwen2-VL-2B backbone, Jina-Reranker-m0 surpasses several larger text-only rerankers, demonstrating clear gains from multimodal information. Furthermore, multimodal models prompted to first reason and then rank (e.g., Qwen2.5-VL-72B, GLM-4.5V, and GPT-5) surpass BGE-Reasoner-Reranker-32B, the best-performing textual reranker specifically trained with reasoning capabilities. GPT-5 achieves the highest overall score, underscoring the importance of utilizing multimodal information with reasoning in tackling the complex retrieval demands posed by MR$^{2}$-Bench.

\vspace{-0.3cm}
\section{Conclusion}
\vspace{-0.3cm}
In this paper, we introduce MR$^2$-Bench, a novel benchmark for the assessment of multimodal reasoning-intensive retrieval. The comprehensive investigation of existing methods reveals that current retrievers perform poorly on MR$^2$-Bench, with the best models achieving only 30.68 nDCG@10. Our experimental results underscore the importance of multimodal information and reasoning capabilities for effectively addressing MR$^2$-Bench, highlighting significant potential for improvement in this research area. Additionally, we demonstrate that techniques such as query rewriting and reranking can enhance performance on MR$^2$-Bench. We anticipate that this benchmark will facilitate future research in multimodal retrieval, contributing to more realistic and challenging AI applications.

\bibliography{iclr2026_conference}
\bibliographystyle{iclr2026_conference}

\appendix

\section*{Appendix}
\section{Use of LLMs}
In preparing this manuscript, large language models (LLMs) were utilized solely for English grammar checking and polishing. All substantive content and analyses were developed independently by the authors. For dataset construction, GPT-5~\citep{GPT-5} was employed only for preliminary filtering of candidate data and generating some challenging negative examples, with all final selections and included negative examples thoroughly reviewed and validated by human experts. The relevant procedures are detailed in the appropriate sections of the paper.

\section{Detailed overview of MR${^2}$-Bench}
\label{appendix:detail-overview}
We provide detailed modalities of queries and documents, along with the instructions for each sub-task in~\Cref{tab:detailed-stats}.
\sisetup{group-separator={,}} 
\newcolumntype{Y}{>{\raggedright\arraybackslash}X}     
\newcolumntype{L}[1]{>{\raggedright\arraybackslash}p{#1}}
\newcolumntype{C}[1]{>{\centering\arraybackslash}m{#1}}

\begin{table}[htbp]
\centering
\setlength{\tabcolsep}{4pt}      
\renewcommand{\arraystretch}{1.08} 
\scriptsize
\begin{tabularx}{\textwidth}{@{}C{2cm} L{1.5cm} L{2.0cm} S[table-format=3] S[table-format=4,group-digits=integer,group-minimum-digits=3,group-separator = {,}] Y@{}}
\toprule
\makecell[c]{\textbf{Meta-Task}} & \textbf{Sub-Task} &
\makecell[l]{\textbf{Modality} $(q \!\to\! c)$} &
\textbf{\#Queries} & \textbf{\#Corpus} & \textbf{Instruction} \\
\midrule
\addlinespace[3pt]
\multirow[c]{6}{*}{\textsc{\makecell[c]{\\ \\ \\ \\ Multimodal\\Knowledge\\Retrieval}}}
& Biology      & $q_{i+t} \!\to\! c_{i/t/i+t}$ & 79  & 4455 & \textit{Find paragraph(s) that could support answering this question.} \\
& Cooking      & $q_{i+t} \!\to\! c_{i/t/i+t}$ & 76  & 2786 & \textit{Find paragraph(s) that could support answering this question.} \\
& Gardening    & $q_{i+t} \!\to\! c_{i/t/i+t}$ & 129 & 5636 & \textit{Find paragraph(s) that could support answering this question.} \\
& Physics      & $q_{i+t} \!\to\! c_{i/t/i+t}$ & 76  & 6656 & \textit{Find paragraph(s) that could support answering this question.} \\
& Chemistry    & $q_{i+t} \!\to\! c_{i/t/i+t}$ & 124 & 4317 & \textit{Find paragraph(s) that could support answering this question.} \\
& EarthScience & $q_{i+t} \!\to\! c_{i/t/i+t}$ & 99  & 3014 & \textit{Find paragraph(s) that could support answering this question.} \\
\cmidrule(lr){1-6}

\addlinespace[1pt]
\multirow[c]{3}{*}{\textsc{\makecell[c]{\\ \\ Visual \\ Illustration\\ Search}}}
& Economics    & $q_t \!\to\! c_i$   & 84  & 7572 & \textit{Find the chart that best supports answering this question.} \\
& Mathematics & $q_t \!\to\! c_i$   & 86  & 944  & \textit{Find the visual proof that best demonstrates this formula.} \\
& Nature  & $q_t \!\to\! c_i$   & 100 & 2017 & \textit{Given a natural-world expert query, find the most relevant image.} \\
\addlinespace[1pt]
\cmidrule(lr){1-6}

\multirow[c]{3}{*}{\textsc{\makecell[c]{\\ \\ \\ Visual \\ Relation}}}
& Spatial  & $q_{i+t} \!\to\! c_i$ & 149 & 1000 & \textit{Given a reference image and a text modification, retrieve the image that best matches the modified reference.} \\
& Visual Puzzle   & $q_i \!\to\! c_i$     & 160 & 5375 & \textit{From a 3×3 grid with one missing cell, retrieve the best candidate image to complete the bottom-right cell based on patterns and relations.} \\
& Analogy & $q_i \!\to\! c_i$     & 147 & 3970 & \textit{Given three images, complete the analogy by retrieving the candidate that applies to the third image the relation from the first to the second.} \\
\bottomrule
\end{tabularx}
\caption{\textbf{The overview of MR$^{2}$-Bench.} MR$^{2}$-Bench consists of three meta-tasks and twelve sub-tasks, totaling 1,309 queries. Subscripts indicate the modalities of the query $q$ and candidate $c$: $i$ denotes image, $t$ denotes text, and $i{+}t$ denotes interleaved image-text.
}
\label{tab:detailed-stats}
\end{table}

\section{More Details of Data Construction for Multimodal Knowledge Retrieval Tasks}
\label{appendix:data-construction}

We collected real posts from the Stack Exchange platform to construct our multimodal knowledge retrieval sub-tasks. Queries are derived from actual user questions, while positive documents are sourced from external links in highly voted answers. We utilize BRIGHT's definition to identify a query's positive document: \textit{A document is relevant only if cited in a highly voted answer and confirmed by annotators and domain experts as aiding in reasoning through the query with critical concepts or theories}~\citep{bright2025}. Given the multimodal nature of the task in MR$^2$-Bench, our annotation process diverges from BRIGHT's construction methodology. The specific steps of our process are summarized as follows:

\textbf{Initial Posts Collection and Filtering.}  
We initiated the process by gathering a substantial set of posts from Stack Exchange. To ensure data quality and relevance, we retained posts meeting specific criteria: (1) the question must contain image(s) essential for understanding the query; (2) the post must have received at least five community votes, indicating reliability; and (3) the answer must include at least one external link to facilitate further content acquisition.

\textbf{Web Page Acquisition and Paragraph Annotation.}  
For each qualifying post, annotators are required to visit the external links provided in the answers and copy the interleaved text-image content in the order it appears, excluding Wikipedia.\footnote{Wikipedia content was automatically extracted using \href{https://playwright.dev/python/}{Playwright} to minimize manual effort.} They then segment this content into paragraphs, preserving images to maintain multimodal information. This process generates a collection of candidate paragraphs for each query, including both text-only and image-containing segments. Initial identification of positive paragraphs is performed using GPT-5~\citep{GPT-5}, followed by expert validation to ensure accuracy and relevance. Only queries with at least one confirmed positive paragraph are included in the final dataset.

\textbf{Incorporation of Challenging Negative Examples.}  
To rigorously assess the reasoning capabilities of evaluation methods, we introduced challenging negative samples for each retained query using two strategies: (1) retrieving topic-related documents from an internal corpus using the query's keywords, with GPT-5 initially verifying they are not false negatives; and (2) using GPT-5 to generate documents that, while topically related, provide unhelpful information. All negative samples were subsequently reviewed by human experts to ensure the integrity of the benchmark.

\section{More Details of Baselines}
\label{appendix:baselines}
In our evaluation, we classify the retriever baseline into two main categories: text embedding models and multimodal embedding models. We assess the Seed1.6-Embedding model~\citep{seed1-6-embedding} via its official API, whereas all other models are evaluated using their publicly available code and open-source checkpoints. Below, we provide a comprehensive overview of the implementation details for all baselines used in the evaluation process.

\subsection{Text Embedding Models}

The evaluated text retrievers include: BGE-M3~\citep{bge-m3}, Qwen3-Embedding~\citep{qwen3-embedding}, ReasonIR~\citep{shao2025reasonir}, BGR-Reasoner-Embed\footnote{\url{https://huggingface.co/BAAI/bge-reasoner-embed-qwen3-8b-0923}}, and Diver-Embed~\citep{diver}. Notably, the last three models have been fine-tuned specifically for reasoning-intensive retrieval tasks, as detailed in their technical reports or repository descriptions.

We consider two input configurations for all text-only retrievers. The first configuration ignores images, utilizing only the textual content from queries and documents; this setup is not applicable to some sub-tasks where either the query or candidates are purely visual. The second configuration employs a caption-augmented approach, where every image in both queries and documents is replaced with a textual description. Specifically, we use the Qwen2.5-VL-7B model~\citep{Qwen2.5-VL} to generate captions for the images with the prompt: \textit{Write a detailed English caption for this image, covering the main objects, their attributes, relationships, actions, layout, and background elements.} Each image in the original input is then substituted with a caption prefixed by its identifier, formatted as \texttt{[IMAGE\_id]: image\_caption}.

\subsection{Multimodal Embedding Models}

The evaluated multimodal retrievers include CLIP~\citep{clip-radford2021learning}, BGE-VL~\citep{megapairs}, GME~\citep{gme2025}, VLM2Vec-V2~\citep{vlm2vec-v2}, MM-Embed~\citep{nv-mm-embed2024}, and Seed1.6-Embedding~\citep{seed1-6-embedding}. All these models can process individual images and texts directly. However, for interleaved image-text data with multiple images, different models require specific handling approaches:

For the CLIP model, we employ a score fusion strategy, following previous work \citep{mbeir-wei2023uniir}. This involves separately embedding the image and text data and then combining these embeddings through element-wise addition to achieve the final image-text representation.

For models that can only input a single image in image-text data, specifically BGE-VL~\citep{megapairs} and MM-Embed~\citep{nv-mm-embed2024}, we create a composite image by tiling multiple images together, which is then processed jointly with the text.

For other models capable of handling interleaved image-text data with multiple images, we preserve the sequence of images and text, allowing their processors to generate interleaved image-text tokens, which are then used to derive the final embeddings.
s and diagnostic analyses are provided in the Appendix.

\section{Detailed Evaluation Metrics of MR$^2$-Bench}
\label{appendix:detailed-evaluation}

In this section, we provide more detailed evaluation results of the embedding models on MR²-Bench. ~\Cref{tab:detailed-main-results-recall@1}, ~\Cref{tab:detailed-main-results-recall@5}, ~\Cref{tab:detailed-main-results-recall@10}, ~\Cref{tab:detailed-main-results-nDCG@5}, and ~\Cref{tab:detailed-main-results-nDCG@20} present the performance of the embedding models in terms of Recall@1, Recall@5, Recall@10, nDCG@5 and nDCG@20.

\definecolor{caprow}{HTML}{F2F2F2}
\begin{table}[htbp]
\begin{adjustbox}{max width=\textwidth}
\renewcommand{\arraystretch}{1.10}   
\setlength{\extrarowheight}{0.2em}
\begin{tabular}{@{}cccccccccccccc@{}}
\toprule
\multicolumn{1}{c|}{\multirow{2}{*}{\textbf{Methods}}} & 
\multicolumn{6}{c|}{\textbf{Multimodal Knowledge Retrieval}} & 
\multicolumn{3}{c|}{\textbf{Visual Illustration}} & 
\multicolumn{3}{c|}{\textbf{Visual Relation}} & 
\multicolumn{1}{l}{\multirow{2}{*}{\textbf{Avg.}}} \\ 
\cmidrule(lr){2-7} \cmidrule(lr){8-10} \cmidrule(lr){11-13}
\multicolumn{1}{c|}{} & \textbf{Bio.} & \textbf{Cook.} & \textbf{Gar.} & \textbf{Phy.} & \textbf{Chem.} & \multicolumn{1}{c|}{\textbf{Earth.}} & 
\textbf{Econ.} & \textbf{Math.} & \multicolumn{1}{c|}{\textbf{Nat.}} & 
\textbf{Spa.} & \textbf{Puzz.} & \multicolumn{1}{c|}{\textbf{Ana.}} & 
\multicolumn{1}{l}{} \\ 
\midrule
\multicolumn{14}{c}{\textcolor{gray}{\textbf{\textit{Text Embedding Models}}}} \\ \midrule
\multicolumn{1}{l|}{BGE-M3} & 3.61 & 2.25 & 3.26 & 2.70 & 1.04 & \multicolumn{1}{c|}{3.77} & - & - & \multicolumn{1}{c|}{-} & - & - & \multicolumn{1}{c|}{-} & - \\
\rowcolor{caprow}
\multicolumn{1}{r|}{\textit{+ Captions}}& 10.22 & 3.23 & 6.44 & 5.00 & 1.43 & \multicolumn{1}{c|}{7.26} & 32.14 & 3.49 & \multicolumn{1}{c|}{6.67} & 4.00 & 0.00 & \multicolumn{1}{c|}{1.36} & 6.77 \\
\addlinespace[2pt]
\multicolumn{1}{l|}{Qwen3} & 5.46  & 2.67 & 3.11 & 1.86 & 1.13 & \multicolumn{1}{c|}{4.67} & - & - & \multicolumn{1}{c|}{-} & - & - & \multicolumn{1}{c|}{-} & - \\
\rowcolor{caprow}
\multicolumn{1}{r|}{\textit{+ Captions}} & 7.21 & 6.87 & 5.15 & 4.84 & 1.20 & \multicolumn{1}{c|}{5.93} & 32.14 & 6.98 & \multicolumn{1}{c|}{4.92} & 4.03 & 0.00 & \multicolumn{1}{c|}{0.68} & 6.66 \\
\addlinespace[2pt]
\multicolumn{1}{l|}{Diver-Emb.} & 5.73 & 2.55 & 4.74 & 1.60 & 0.38 & \multicolumn{1}{c|}{3.71} & - & - & \multicolumn{1}{c|}{-} & - & - & \multicolumn{1}{c|}{-} & - \\
\rowcolor{caprow}
\multicolumn{1}{r|}{\textit{+ Captions}} & 12.37 & 7.06 & 9.87 & 4.60 & 2.39 & \multicolumn{1}{c|}{6.60} & 36.90 & 8.14 & \multicolumn{1}{c|}{3.00} & 3.33 & 0.00 & \multicolumn{1}{c|}{0.68} & 7.91 \\
\addlinespace[2pt]
\multicolumn{1}{l|}{BGE-Rea.} & 3.69 & 3.13 & 4.10 & 2.59 & 1.36 & \multicolumn{1}{c|}{4.64} & - & - & \multicolumn{1}{c|}{-} & - & - & \multicolumn{1}{c|}{-} & - \\
\rowcolor{caprow}
\multicolumn{1}{r|}{\textit{+ Captions}} & 16.03 & 9.84 & 9.74 & 6.19 & 1.24 & \multicolumn{1}{c|}{10.28} & 41.67 & 12.21 & \multicolumn{1}{c|}{1.67} & 2.00 & 0.00 & \multicolumn{1}{c|}{0.68} & 9.29 \\
\addlinespace[2pt]
\multicolumn{1}{l|}{ReasonIR} & 7.68 & 3.13 & 3.75 & 4.35 & 0.91 & \multicolumn{1}{c|}{4.21} & - & - & \multicolumn{1}{c|}{-} & - & - & \multicolumn{1}{c|}{-} & - \\
\rowcolor{caprow}
\multicolumn{1}{r|}{\textit{+ Captions}} & 16.87 & 13.81 & 7.13 & 5.32 & 3.50 & \multicolumn{1}{c|}{11.59} & 39.29 & 7.56 & \multicolumn{1}{c|}{6.58} & 2.00 & 0.00 & \multicolumn{1}{c|}{0.68} & 9.53 \\
 \midrule
\multicolumn{14}{c}{\textcolor{gray}{\textbf{\textit{Multimodal Embedding Models}}}} \\ \midrule
\multicolumn{1}{l|}{CLIP} & 12.49 & 8.28 & 4.37 & 2.58 & 1.42 & \multicolumn{1}{c|}{11.72} & 3.57 & 1.16 & \multicolumn{1}{c|}{10.92} & 12.67 & 0.00 & \multicolumn{1}{c|}{0.00} & 5.77 \\
\multicolumn{1}{l|}{BGE-VL} & 8.96 & 2.30 & 2.93 & 4.35 & 0.32 & \multicolumn{1}{c|}{4.81} & 34.52 & 6.98 & \multicolumn{1}{c|}{10.83} & 2.01 & 0.00 & \multicolumn{1}{c|}{1.36} & 6.62 \\
\multicolumn{1}{l|}{GME} & 10.07 & 14.48 & 7.84 & 3.97 & 1.43 & \multicolumn{1}{c|}{8.39} & 21.43 & 2.33 & \multicolumn{1}{c|}{8.08} & 8.00 & 0.00 & \multicolumn{1}{c|}{3.40} & 7.45 \\
\multicolumn{1}{l|}{VLM2Vec} & 13.58 & 13.73 & 5.41 & 3.73 & 1.44 & \multicolumn{1}{c|}{14.54} & 38.10 & 3.49 & \multicolumn{1}{c|}{9.53} & 4.00 & 0.62 & \multicolumn{1}{c|}{0.68} & 9.07 \\
\multicolumn{1}{l|}{MM-Emb.} & 17.18 & 20.81 & 7.10 & 7.05 & 4.35 & \multicolumn{1}{c|}{17.54} & 34.52 & 9.59 & \multicolumn{1}{c|}{11.25} & 11.33 & 0.00 & \multicolumn{1}{c|}{0.00} & 11.73 \\
\multicolumn{1}{l|}{Seed-1.6} & 13.65 & 9.02 & 9.85 & 5.20 & 3.69 & \multicolumn{1}{c|}{9.81} & 33.33 & 6.98 & \multicolumn{1}{c|}{19.33} & 8.00 & 0.00 & \multicolumn{1}{c|}{0.00} & 9.91 \\
\bottomrule
\end{tabular}
\end{adjustbox}
\caption{{The overall performance of embedding models on MR$^{2}$-Bench in terms of the recall@1.}}
\label{tab:detailed-main-results-recall@1}
\end{table}

\definecolor{caprow}{HTML}{F2F2F2}
\begin{table}[htbp]
\begin{adjustbox}{max width=\textwidth}
\renewcommand{\arraystretch}{1.1}   
\setlength{\extrarowheight}{0.2em}
\begin{tabular}{@{}cccccccccccccc@{}}
\toprule
\multicolumn{1}{c|}{\multirow{2}{*}{\textbf{Methods}}} & 
\multicolumn{6}{c|}{\textbf{Multimodal Knowledge Retrieval}} & 
\multicolumn{3}{c|}{\textbf{Visual Illustration}} & 
\multicolumn{3}{c|}{\textbf{Visual Relation}} & 
\multicolumn{1}{l}{\multirow{2}{*}{\textbf{Avg.}}} \\ 
\cmidrule(lr){2-7} \cmidrule(lr){8-10} \cmidrule(lr){11-13}
\multicolumn{1}{c|}{} & \textbf{Bio.} & \textbf{Cook.} & \textbf{Gar.} & \textbf{Phy.} & \textbf{Chem.} & \multicolumn{1}{c|}{\textbf{Earth.}} & 
\textbf{Econ.} & \textbf{Math.} & \multicolumn{1}{c|}{\textbf{Nat.}} & 
\textbf{Spa.} & \textbf{Puzz.} & \multicolumn{1}{c|}{\textbf{Ana.}} & 
\multicolumn{1}{l}{} \\ 
\midrule
\multicolumn{14}{c}{\textcolor{gray}{\textbf{\textit{Text Embedding Models}}}} \\ \midrule
\multicolumn{1}{l|}{BGE-M3} & 14.09 & 10.43 & 11.66 & 10.12 & 7.27 & \multicolumn{1}{c|}{12.78} & - & - & \multicolumn{1}{c|}{-} & - & - & \multicolumn{1}{c|}{-} & - \\
\rowcolor{caprow}
\multicolumn{1}{r|}{\textit{+ Captions}} & 28.85 & 23.21 & 17.32 & 13.60 & 8.76 & \multicolumn{1}{c|}{20.66} & 53.57 & 10.85 & \multicolumn{1}{c|}{20.25} & 11.33 & 0.00 & \multicolumn{1}{c|}{3.40} & 17.65 \\
\addlinespace[2pt]
\multicolumn{1}{l|}{Qwen3} & 17.36 & 12.48 & 12.17 & 11.93 & 6.38 & \multicolumn{1}{c|}{15.12} & - & - &  \multicolumn{1}{c|}{-} & - & - & \multicolumn{1}{c|}{-} & - \\
\rowcolor{caprow}
\multicolumn{1}{r|}{\textit{+ Captions}} & 24.76 & 27.10 & 16.30 & 13.73 & 6.61 & \multicolumn{1}{c|}{17.95} & 60.71 & 30.33 & \multicolumn{1}{c|}{24.42} & 11.41 & 0.00 & \multicolumn{1}{c|}{5.44} & 19.90 \\
\addlinespace[2pt]
\multicolumn{1}{l|}{Diver-Emb.} & 24.54 & 12.43 & 15.95 & 13.61 & 8.07 & \multicolumn{1}{c|}{19.33} & - & - & \multicolumn{1}{c|}{-} & - & - & \multicolumn{1}{c|}{-} & - \\
\rowcolor{caprow}
\multicolumn{1}{r|}{\textit{+ Captions}} & 30.82 & 27.00 & 20.75 & 16.47 & 11.35 & \multicolumn{1}{c|}{29.76} & 65.48 & 37.50 & \multicolumn{1}{c|}{21.67} & 10.00 & 0.00 & \multicolumn{1}{c|}{10.88} & 23.47 \\
\addlinespace[2pt]
\multicolumn{1}{l|}{BGE-Rea.} & 23.36 & 8.80 & 15.83 & 13.79 & 8.05 & \multicolumn{1}{c|}{19.01} & - & - & \multicolumn{1}{c|}{-} & - & - & \multicolumn{1}{c|}{-} & - \\
\rowcolor{caprow}
\multicolumn{1}{r|}{\textit{+ Captions}} & 33.49 & 32.50 & 25.09 & 17.00 & 12.56 & \multicolumn{1}{c|}{26.61} & 70.24 & 46.71 & \multicolumn{1}{c|}{25.42} & 8.67 & 0.00 & \multicolumn{1}{c|}{6.80} & 25.42 \\
\addlinespace[2pt]
\multicolumn{1}{l|}{ReasonIR} & 26.10 & 16.55 & 14.73 & 14.94 & 10.08 & \multicolumn{1}{c|}{20.38} & - & - & \multicolumn{1}{c|}{-} & - & - & \multicolumn{1}{c|}{-} & - \\
\rowcolor{caprow}
\multicolumn{1}{r|}{\textit{+ Captions}} & 33.01 & 36.37 & 16.49 & 20.05 & 17.45 & \multicolumn{1}{c|}{36.66} & 61.90 & 20.93 & \multicolumn{1}{c|}{24.33} & 6.00 & 0.00 & \multicolumn{1}{c|}{8.16} & 23.45 \\
 \midrule
\multicolumn{14}{c}{\textcolor{gray}{\textbf{\textit{Multimodal Embedding Models}}}} \\ \midrule
\multicolumn{1}{l|}{CLIP} & 27.54 & 28.63 & 9.72 & 7.60 & 4.01 & \multicolumn{1}{c|}{29.62} & 16.67 & 4.65 & \multicolumn{1}{c|}{48.17} & 22.67 & 0.00 & \multicolumn{1}{c|}{6.80} & 17.17 \\
\multicolumn{1}{l|}{BGE-VL} & 22.17 & 12.55 & 10.96 & 15.18 & 6.30 & \multicolumn{1}{c|}{15.22} & 63.10 & 17.64 & \multicolumn{1}{c|}{47.33} & 10.07 & 0.00 & \multicolumn{1}{c|}{5.44} & 18.83 \\
\multicolumn{1}{l|}{GME} & 27.06 & 33.94 & 15.53 & 13.36 & 6.25 & \multicolumn{1}{c|}{24.78} & 45.24 & 6.40 & \multicolumn{1}{c|}{37.42} & 20.67 & 0.00 & \multicolumn{1}{c|}{12.24} & 20.24 \\
\multicolumn{1}{l|}{VLM2Vec} & 30.64 & 34.61 & 18.89 & 12.44 & 7.36 & \multicolumn{1}{c|}{32.18} & 55.95 & 17.25 & \multicolumn{1}{c|}{31.75} & 17.33 & 0.63 & \multicolumn{1}{c|}{4.08} & 21.93 \\
\multicolumn{1}{l|}{MM-Emb.} & 38.24 & 48.89 & 21.07 & 21.03 & 16.53 & \multicolumn{1}{c|}{42.79} & 48.81 & 22.58 & \multicolumn{1}{c|}{42.08} & 28.00 & 0.00 & \multicolumn{1}{c|}{6.12} & 28.01 \\
\multicolumn{1}{l|}{Seed-1.6} & 31.93 & 32.51 & 28.95 & 22.17 & 14.52 & \multicolumn{1}{c|}{31.65} & 69.05 & 38.76 & \multicolumn{1}{c|}{61.25} & 19.33 & 0.63 & \multicolumn{1}{c|}{8.16} & 29.91 \\
\bottomrule
\end{tabular}
\end{adjustbox}
\caption{{The overall performance of embedding models on MR$^{2}$-Bench in terms of the recall@5.}}
\vspace{0.7cm}
\label{tab:detailed-main-results-recall@5}
\end{table}

\definecolor{caprow}{HTML}{F2F2F2}
\begin{table}[htbp]
\begin{adjustbox}{max width=\textwidth}
\renewcommand{\arraystretch}{1.1}   
\setlength{\extrarowheight}{0.2em}
\begin{tabular}{@{}cccccccccccccc@{}}
\toprule
\multicolumn{1}{c|}{\multirow{2}{*}{\textbf{Methods}}} & 
\multicolumn{6}{c|}{\textbf{Multimodal Knowledge Retrieval}} & 
\multicolumn{3}{c|}{\textbf{Visual Illustration}} & 
\multicolumn{3}{c|}{\textbf{Visual Relation}} & 
\multicolumn{1}{l}{\multirow{2}{*}{\textbf{Avg.}}} \\ 
\cmidrule(lr){2-7} \cmidrule(lr){8-10} \cmidrule(lr){11-13}
\multicolumn{1}{c|}{} & \textbf{Bio.} & \textbf{Cook.} & \textbf{Gar.} & \textbf{Phy.} & \textbf{Chem.} & \multicolumn{1}{c|}{\textbf{Earth.}} & 
\textbf{Econ.} & \textbf{Math.} & \multicolumn{1}{c|}{\textbf{Nat.}} & 
\textbf{Spa.} & \textbf{Puzz.} & \multicolumn{1}{c|}{\textbf{Ana.}} & 
\multicolumn{1}{l}{} \\ 
\midrule
\multicolumn{14}{c}{\textcolor{gray}{\textbf{\textit{Text Embedding Models}}}} \\ \midrule
\multicolumn{1}{l|}{BGE-M3} & 25.92 & 18.48 & 17.29 & 17.00 & 9.11 & \multicolumn{1}{c|}{22.31} & - & - & \multicolumn{1}{c|}{-} & - & - & \multicolumn{1}{c|}{-} & - \\
\rowcolor{caprow}
\multicolumn{1}{r|}{\textit{+ Captions}} & 39.67 & 35.42 & 20.81 & 21.23 & 14.02 & \multicolumn{1}{c|}{32.55} & 61.90 & 19.57 & \multicolumn{1}{c|}{33.83} & 16.67 & 0.00 & \multicolumn{1}{c|}{7.48} & 25.26 \\
\addlinespace[2pt]
\multicolumn{1}{l|}{Qwen3} & 32.83 & 30.81 & 17.08 & 20.64 & 13.94 & \multicolumn{1}{c|}{27.05} & - & - & \multicolumn{1}{c|}{-} & - & - & \multicolumn{1}{c|}{-} & - \\
\rowcolor{caprow}
\multicolumn{1}{r|}{\textit{+ Captions}} & 33.91 & 38.52 & 24.39 & 21.03 & 15.41 & \multicolumn{1}{c|}{31.27} & 67.86 & 38.08 & \multicolumn{1}{c|}{39.67} & 16.11 & 0.00 & \multicolumn{1}{c|}{8.84} & 27.92 \\
\addlinespace[2pt]
\multicolumn{1}{l|}{Diver-Emb.} & 35.18 & 25.13 & 20.01 & 22.07 & 16.42 & \multicolumn{1}{c|}{33.76} & - & - & \multicolumn{1}{c|}{-} & - & - & \multicolumn{1}{c|}{-} & - \\
\rowcolor{caprow}
\multicolumn{1}{r|}{\textit{+ Captions}} & 43.81 & 42.45 & 26.21 & 25.29 & 22.66 & \multicolumn{1}{c|}{43.71} & 70.24 & 43.31 & \multicolumn{1}{c|}{40.83} & 16.00 & 0.00 & \multicolumn{1}{c|}{15.65} & 32.51 \\
\addlinespace[2pt]
\multicolumn{1}{l|}{BGE-Rea.} & 41.29 & 24.18 & 23.80 & 23.82 & 17.24 & \multicolumn{1}{c|}{37.44} & - & - & \multicolumn{1}{c|}{-} & - & - & \multicolumn{1}{c|}{-} & - \\
\rowcolor{caprow}
\multicolumn{1}{r|}{\textit{+ Captions}} & 46.35 & 42.84 & 29.78 & 24.51 & 22.96 & \multicolumn{1}{c|}{43.02} & 79.76 & 58.53 & \multicolumn{1}{c|}{40.42} & 12.00 & 0.00 & \multicolumn{1}{c|}{11.56} & 34.31 \\
\addlinespace[2pt]
\multicolumn{1}{l|}{ReasonIR} & 37.42 & 29.45 & 22.93 & 24.88 & 16.21 & \multicolumn{1}{c|}{34.28} & - & - & \multicolumn{1}{c|}{-} & - & - & \multicolumn{1}{c|}{-} & - \\
\rowcolor{caprow}
\multicolumn{1}{r|}{\textit{+ Captions}} & 46.05 & 50.89 & 20.99 & 28.93 & 25.72 & \multicolumn{1}{c|}{52.31} & 69.05 & 28.88 & \multicolumn{1}{c|}{44.08} & 10.00 & 0.00 & \multicolumn{1}{c|}{12.93} & 32.48 \\
 \midrule
\multicolumn{14}{c}{\textcolor{gray}{\textbf{\textit{Multimodal Embedding Models}}}} \\ \midrule
\multicolumn{1}{l|}{CLIP} & 33.08 & 38.08 & 14.38 & 14.05 & 5.64 & \multicolumn{1}{c|}{38.85} & 26.19 & 12.21 & \multicolumn{1}{c|}{70.42} & 31.33 & 0.63 & \multicolumn{1}{c|}{11.56} & 24.70 \\
\multicolumn{1}{l|}{BGE-VL} & 38.03 & 27.68 & 16.66 & 22.85 & 11.62 & \multicolumn{1}{c|}{28.46} & 66.67 & 23.45 & \multicolumn{1}{c|}{67.83} & 12.08 & 0.00 & \multicolumn{1}{c|}{12.93} & 27.35 \\
\multicolumn{1}{l|}{GME} & 35.13 & 45.64 & 21.93 & 19.66 & 13.14 & \multicolumn{1}{c|}{36.10} & 54.76 & 15.50 & \multicolumn{1}{c|}{57.17} & 25.33 & 0.63 & \multicolumn{1}{c|}{23.13} & 29.01 \\
\multicolumn{1}{l|}{VLM2Vec} & 41.02 & 44.31 & 23.69 & 20.66 & 13.20 & \multicolumn{1}{c|}{39.49} & 66.67 & 27.23 & \multicolumn{1}{c|}{47.67} & 27.33 & 0.63 & \multicolumn{1}{c|}{14.97} & 30.57 \\
\multicolumn{1}{l|}{MM-Emb.} & 50.98 & 55.18 & 26.60 & 28.91 & 22.79 & \multicolumn{1}{c|}{54.61} & 51.19 & 35.08 & \multicolumn{1}{c|}{68.42} & 35.33 & 0.63 & \multicolumn{1}{c|}{14.29} & 37.00 \\
\multicolumn{1}{l|}{Seed-1.6} & 47.99 & 49.13 & 38.60 & 30.05 & 26.32 & \multicolumn{1}{c|}{48.90} & 79.76 & 47.87 & \multicolumn{1}{c|}{84.17} & 30.67 & 2.50 & \multicolumn{1}{c|}{22.45} & 42.37 \\
\bottomrule
\end{tabular}
\end{adjustbox}
\caption{{The overall performance of embedding models on MR$^{2}$-Bench in terms of the recall@10.}}
\label{tab:detailed-main-results-recall@10}
\end{table}

\definecolor{caprow}{HTML}{F2F2F2}
\begin{table}[htbp]
\begin{adjustbox}{max width=\textwidth}
\renewcommand{\arraystretch}{1.1}   
\setlength{\extrarowheight}{0.2em}
\begin{tabular}{@{}cccccccccccccc@{}}
\toprule
\multicolumn{1}{c|}{\multirow{2}{*}{\textbf{Methods}}} & 
\multicolumn{6}{c|}{\textbf{Multimodal Knowledge Retrieval}} & 
\multicolumn{3}{c|}{\textbf{Visual Illustration}} & 
\multicolumn{3}{c|}{\textbf{Visual Relation}} & 
\multicolumn{1}{l}{\multirow{2}{*}{\textbf{Avg.}}} \\ 
\cmidrule(lr){2-7} \cmidrule(lr){8-10} \cmidrule(lr){11-13}
\multicolumn{1}{c|}{} & \textbf{Bio.} & \textbf{Cook.} & \textbf{Gar.} & \textbf{Phy.} & \textbf{Chem.} & \multicolumn{1}{c|}{\textbf{Earth.}} & 
\textbf{Econ.} & \textbf{Math.} & \multicolumn{1}{c|}{\textbf{Nat.}} & 
\textbf{Spa.} & \textbf{Puzz.} & \multicolumn{1}{c|}{\textbf{Ana.}} & 
\multicolumn{1}{l}{} \\ 
\midrule
\multicolumn{14}{c}{\textcolor{gray}{\textbf{\textit{Text Embedding Models}}}} \\ \midrule
\multicolumn{1}{l|}{BGE-M3} & 14.89 & 10.15 & 9.78 & 13.18 & 5.33 & \multicolumn{1}{c|}{13.23} & - & - & \multicolumn{1}{c|}{-} & - & - & \multicolumn{1}{c|}{-} & - \\
\rowcolor{caprow}
\multicolumn{1}{r|}{\textit{+ Captions}} & 32.01 & 19.33 & 17.27 & 21.47 & 8.02 & \multicolumn{1}{c|}{21.21} & 42.82 & 7.13 & \multicolumn{1}{c|}{17.66} & 7.81 & 0.00 & \multicolumn{1}{c|}{2.18} & 16.41 \\
\addlinespace[2pt]
\multicolumn{1}{l|}{Qwen3} & 18.71 & 13.35 & 10.80 & 14.72 & 5.76 & \multicolumn{1}{c|}{15.71} & - & - &\multicolumn{1}{c|}{-} & - & - & \multicolumn{1}{c|}{-} & - \\
\rowcolor{caprow}
\multicolumn{1}{r|}{\textit{+ Captions}} & 27.36 & 24.45 & 15.25 & 20.60 & 6.75 & \multicolumn{1}{c|}{18.77} & 47.01 & 18.56 & \multicolumn{1}{c|}{19.48} & 7.70 & 0.00 & \multicolumn{1}{c|}{3.20} & 17.43 \\
\addlinespace[2pt]
\multicolumn{1}{l|}{Diver-Emb.} & 24.49 & 11.88 & 13.45 & 16.24 & 6.70 & \multicolumn{1}{c|}{17.47} & - & - & \multicolumn{1}{c|}{-} & - & - & \multicolumn{1}{c|}{-} & - \\
\rowcolor{caprow}
\multicolumn{1}{r|}{\textit{+ Captions}} & 36.03 & 24.59 & 20.72 & 21.95 & 10.66 & \multicolumn{1}{c|}{26.57} & 53.13 & 23.94 & \multicolumn{1}{c|}{16.49} & 6.60 & 0.00 & \multicolumn{1}{c|}{5.97} & 20.56 \\
\addlinespace[2pt]
\multicolumn{1}{l|}{BGE-Rea.} & 23.63 & 8.90 & 12.79 & 18.98 & 6.92 & \multicolumn{1}{c|}{19.62} & - & - & \multicolumn{1}{c|}{-} & - & - & \multicolumn{1}{c|}{-} & - \\
\rowcolor{caprow}
\multicolumn{1}{r|}{\textit{+ Captions}} & 40.07 & 30.43 & 23.44 & 25.26 & 10.06 & \multicolumn{1}{c|}{28.38} & 57.44 & 30.40 & \multicolumn{1}{c|}{18.43} & 5.59 & 0.00 & \multicolumn{1}{c|}{4.11} & 22.80 \\
\addlinespace[2pt]
\multicolumn{1}{l|}{ReasonIR} & 26.90 & 14.77 & 13.12 & 19.15 & 7.68 & \multicolumn{1}{c|}{17.96} & - & - & \multicolumn{1}{c|}{-} & - & - & \multicolumn{1}{c|}{-} & - \\
\rowcolor{caprow}
\multicolumn{1}{r|}{\textit{+ Captions}} & 42.83 & 36.92 & 17.98 & 26.60 & 14.74 & \multicolumn{1}{c|}{36.36} & 52.01 & 14.34 & \multicolumn{1}{c|}{21.23} & 4.30 & 0.00 & \multicolumn{1}{c|}{4.36} & 22.64 \\
 \midrule
\multicolumn{14}{c}{\textcolor{gray}{\textbf{\textit{Multimodal Embedding Models}}}} \\ \midrule
\multicolumn{1}{l|}{CLIP} & 33.19 & 27.83 & 10.42 & 15.34 & 4.90 & \multicolumn{1}{c|}{30.50} & 9.91 & 3.13 & \multicolumn{1}{c|}{39.38} & 18.14 & 0.00 & \multicolumn{1}{c|}{3.57} & 16.36 \\
\multicolumn{1}{l|}{BGE-VL} & 26.00 & 12.99 & 9.92 & 18.10 & 5.71 & \multicolumn{1}{c|}{14.74} & 49.60 & 12.44 & \multicolumn{1}{c|}{39.07} & 5.83 & 0.00 & \multicolumn{1}{c|}{3.53} & 16.49 \\
\multicolumn{1}{l|}{GME} & 33.91 & 35.58 & 17.03 & 18.41 & 5.46 & \multicolumn{1}{c|}{25.61} & 33.89 & 4.18 & \multicolumn{1}{c|}{30.64} & 14.21 & 0.00 & \multicolumn{1}{c|}{7.64} & 18.88 \\
\multicolumn{1}{l|}{VLM2Vec} & 38.31 & 36.87 & 18.75 & 19.66 & 7.46 & \multicolumn{1}{c|}{34.05} & 47.87 & 10.86 & \multicolumn{1}{c|}{28.12} & 10.74 & 0.63 & \multicolumn{1}{c|}{2.46} & 21.31 \\
\multicolumn{1}{l|}{MM-Emb.} & 48.80 & 50.58 & 22.22 & 30.84 & 15.50 & \multicolumn{1}{c|}{44.52} & 42.15 & 17.22 & \multicolumn{1}{c|}{37.07} & 20.34 & 0.00 & \multicolumn{1}{c|}{3.29} & 27.71 \\
\multicolumn{1}{l|}{Seed-1.6} & 36.14 & 32.45 & 28.34 & 27.69 & 13.46 & \multicolumn{1}{c|}{31.52} & 52.63 & 22.95 & \multicolumn{1}{c|}{55.12} & 13.67 & 0.31 & \multicolumn{1}{c|}{4.49} & 26.56 \\
\bottomrule
\end{tabular}
\end{adjustbox}
\caption{{The overall performance of embedding models on MR$^{2}$-Bench in terms of the nDCG@5.}}
\label{tab:detailed-main-results-nDCG@5}
\end{table}

\definecolor{caprow}{HTML}{F2F2F2}
\begin{table}[htbp]
\begin{adjustbox}{max width=\textwidth}
\renewcommand{\arraystretch}{1.10}   
\setlength{\extrarowheight}{0.2em}
\begin{tabular}{@{}cccccccccccccc@{}}
\toprule
\multicolumn{1}{c|}{\multirow{2}{*}{\textbf{Methods}}} & 
\multicolumn{6}{c|}{\textbf{Multimodal Knowledge Retrieval}} & 
\multicolumn{3}{c|}{\textbf{Visual Illustration}} & 
\multicolumn{3}{c|}{\textbf{Visual Relation}} & 
\multicolumn{1}{l}{\multirow{2}{*}{\textbf{Avg.}}} \\ 
\cmidrule(lr){2-7} \cmidrule(lr){8-10} \cmidrule(lr){11-13}
\multicolumn{1}{c|}{} & \textbf{Bio.} & \textbf{Cook.} & \textbf{Gar.} & \textbf{Phy.} & \textbf{Chem.} & \multicolumn{1}{c|}{\textbf{Earth.}} & 
\textbf{Econ.} & \textbf{Math.} & \multicolumn{1}{c|}{\textbf{Nat.}} & 
\textbf{Spa.} & \textbf{Puzz.} & \multicolumn{1}{c|}{\textbf{Ana.}} & 
\multicolumn{1}{l}{} \\ 
\midrule
\multicolumn{14}{c}{\textcolor{gray}{\textbf{\textit{Text Embedding Models}}}} \\ \midrule
\multicolumn{1}{l|}{BGE-M3} & 22.66 & 16.33 & 13.67 & 17.72 & 7.89 & \multicolumn{1}{c|}{19.84} & - & - & \multicolumn{1}{c|}{-} & - & - & \multicolumn{1}{c|}{-} & - \\
\rowcolor{caprow}
\multicolumn{1}{r|}{\textit{+ Captions}} & 37.22 & 28.43 & 19.70 & 24.36 & 11.39 & \multicolumn{1}{c|}{30.06} & 47.59 & 12.65 & \multicolumn{1}{c|}{27.18} & 10.95 & 0.00 & \multicolumn{1}{c|}{5.14} & 21.22 \\
\addlinespace[2pt]
\multicolumn{1}{l|}{Qwen3} & 30.34 & 24.25 & 15.39 & 20.39 & 12.16 & \multicolumn{1}{c|}{24.16} & - & - & \multicolumn{1}{c|}{-} & - & - & \multicolumn{1}{c|}{-} & - \\
\rowcolor{caprow}
\multicolumn{1}{r|}{\textit{+ Captions}} & 36.76 & 33.24 & 21.41 & 24.91 & 13.21 & \multicolumn{1}{c|}{29.03} & 49.75 & 24.00 & \multicolumn{1}{c|}{30.48} & 10.96 & 0.00 & \multicolumn{1}{c|}{6.70} & 23.37 \\
\addlinespace[2pt]
\multicolumn{1}{l|}{Diver-Emb.} & 32.45 & 23.00 & 17.59 & 24.00 & 13.65 & \multicolumn{1}{c|}{28.12} & - & - & \multicolumn{1}{c|}{-} & - & - & \multicolumn{1}{c|}{-} & - \\
\rowcolor{caprow}
\multicolumn{1}{r|}{\textit{+ Captions}} & 43.90 & 36.21 & 25.50 & 28.19 & 18.26 & \multicolumn{1}{c|}{37.13} & 58.30 & 29.83 & \multicolumn{1}{c|}{29.68} & 10.52 & 0.17 & \multicolumn{1}{c|}{8.86} & 27.21 \\
\addlinespace[2pt]
\multicolumn{1}{l|}{BGE-Rea.} & 33.43 & 21.87 & 18.97 & 25.40 & 14.43 & \multicolumn{1}{c|}{30.65} & - & - & \multicolumn{1}{c|}{-} & - & - & \multicolumn{1}{c|}{-} & - \\
\rowcolor{caprow}
\multicolumn{1}{r|}{\textit{+ Captions}} & 47.04 & 39.94 & 28.24 & 29.94 & 19.20 & \multicolumn{1}{c|}{40.59} & 63.25 & 35.94 & \multicolumn{1}{c|}{33.36} & 8.87 & 0.00 & \multicolumn{1}{c|}{7.89} & 29.52 \\
\addlinespace[2pt]
\multicolumn{1}{l|}{ReasonIR} & 36.90 & 24.69 & 18.92 & 25.97 & 13.12 & \multicolumn{1}{c|}{30.39} & - & - & \multicolumn{1}{c|}{-} & - & - & \multicolumn{1}{c|}{-} & - \\
\rowcolor{caprow}
\multicolumn{1}{r|}{\textit{+ Captions}} & 48.18 & 45.34 & 21.83 & 28.71 & 21.15 & \multicolumn{1}{c|}{44.74} & 57.61 & 19.33 & \multicolumn{1}{c|}{36.48} & 6.19 & 0.00 & \multicolumn{1}{c|}{8.91} & 28.21 \\
 \midrule
\multicolumn{14}{c}{\textcolor{gray}{\textbf{\textit{Multimodal Embedding Models}}}} \\ \midrule
\multicolumn{1}{l|}{CLIP} & 35.49 & 31.94 & 13.96 & 16.53 & 6.01 & \multicolumn{1}{c|}{34.38} & 14.76 & 6.57 & \multicolumn{1}{c|}{56.32} & 23.04 & 0.53 & \multicolumn{1}{c|}{6.71} & 20.52 \\
\multicolumn{1}{l|}{BGE-VL} & 36.96 & 26.11 & 17.09 & 23.25 & 9.47 & \multicolumn{1}{c|}{26.61} & 52.91 & 16.12 & \multicolumn{1}{c|}{53.97} & 7.71 & 0.17 & \multicolumn{1}{c|}{8.46} & 23.24 \\
\multicolumn{1}{l|}{GME} & 38.17 & 43.48 & 20.72 & 21.56 & 10.82 & \multicolumn{1}{c|}{33.43} & 40.88 & 9.08 & \multicolumn{1}{c|}{45.85} & 18.74 & 0.22 & \multicolumn{1}{c|}{14.88} & 24.82 \\
\multicolumn{1}{l|}{VLM2Vec} & 42.11 & 43.24 & 21.36 & 21.93 & 11.63 & \multicolumn{1}{c|}{38.91} & 55.66 & 18.46 & \multicolumn{1}{c|}{40.38} & 17.18 & 0.63 & \multicolumn{1}{c|}{8.77} & 26.69 \\
\multicolumn{1}{l|}{MM-Emb.} & 51.83 & 54.36 & 26.38 & 32.74 & 20.39 & \multicolumn{1}{c|}{51.77} & 45.99 & 22.91 & \multicolumn{1}{c|}{55.04} & 23.97 & 0.36 & \multicolumn{1}{c|}{8.00} & 32.81 \\
\multicolumn{1}{l|}{Seed-1.6} & 46.01 & 43.31 & 35.86 & 32.99 & 22.85 & \multicolumn{1}{c|}{43.71} & 58.25 & 28.38 & \multicolumn{1}{c|}{69.97} & 21.20 & 1.67 & \multicolumn{1}{c|}{11.76} & 34.66 \\
\bottomrule
\end{tabular}
\end{adjustbox}
\caption{{The overall performance of embedding models on MR$^{2}$-Bench in terms of the nDCG@20.}}
\label{tab:detailed-main-results-nDCG@20}
\end{table}

\section{More Details of Implementation for Query Rewriting}
\label{appendix:query-rewriting}
\begin{figure}[htbp]
    \centering
    \includegraphics[width=\linewidth]{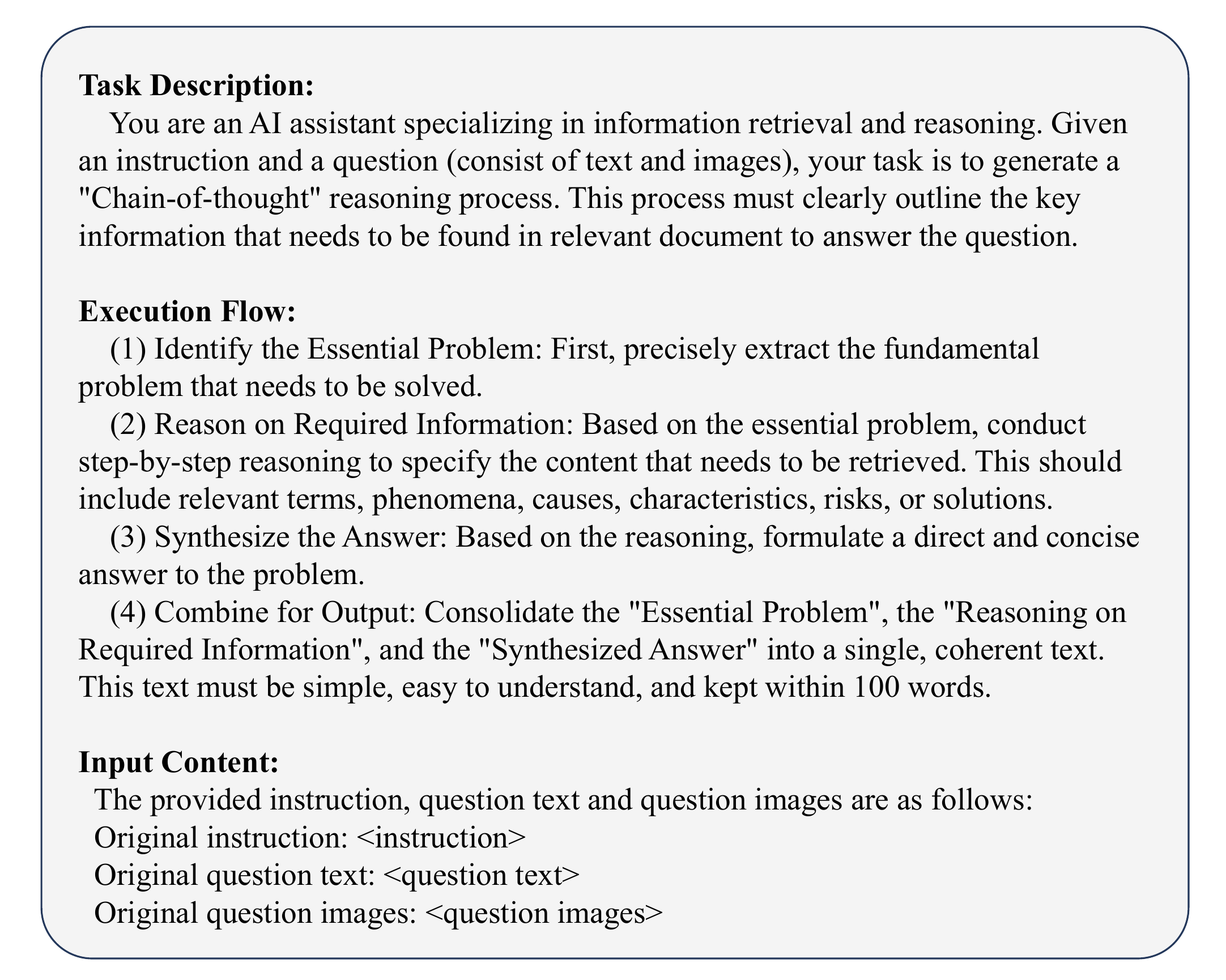}
    \caption{Prompt used by GPT-5 for query rewriting.}
    \label{fig:prompt_query_rewrite}
\end{figure}

Given the strong reasoning capabilities of Multimodal Large Language Models (MLLMs), we take advantage of their ability to produce explicit step-by-step chain-of-thought reasoning in order to improve the effectiveness of query rewriting and thereby enhance retrieval performance.
Instead of relying on a single direct reformulation, we design a prompting strategy that guides the MLLM through a structured reasoning process. Concretely, the model is first asked to (i) identify the most salient subquestions that are implicitly contained in the given instruction and query, ensuring that complex or multifaceted information needs are decomposed into clear components. Next, the model is prompted to (ii) reason step-by-step about what types of evidence, textual patterns, and document attributes would be necessary for relevant sources to contain, which encourages a more targeted and discriminative retrieval process. Finally, model (iii) produces both an explicit reasoning trace, which captures its internal deliberation, and a set of candidate rewritten queries or answers that can be used to drive retrieval more effectively. 
We employ GPT-5~\citep{GPT-5}, the SOTA multimodal reasoning model, to perform query rewriting. The prompt is provided in~\Cref{fig:prompt_query_rewrite}.

\section{More Details of Reranking}
\subsection{IMPLEMENTATION DETAILS }
\label{appendix: rerank settings}

\begin{figure}[htbp]
    \centering
    \includegraphics[width=\linewidth]{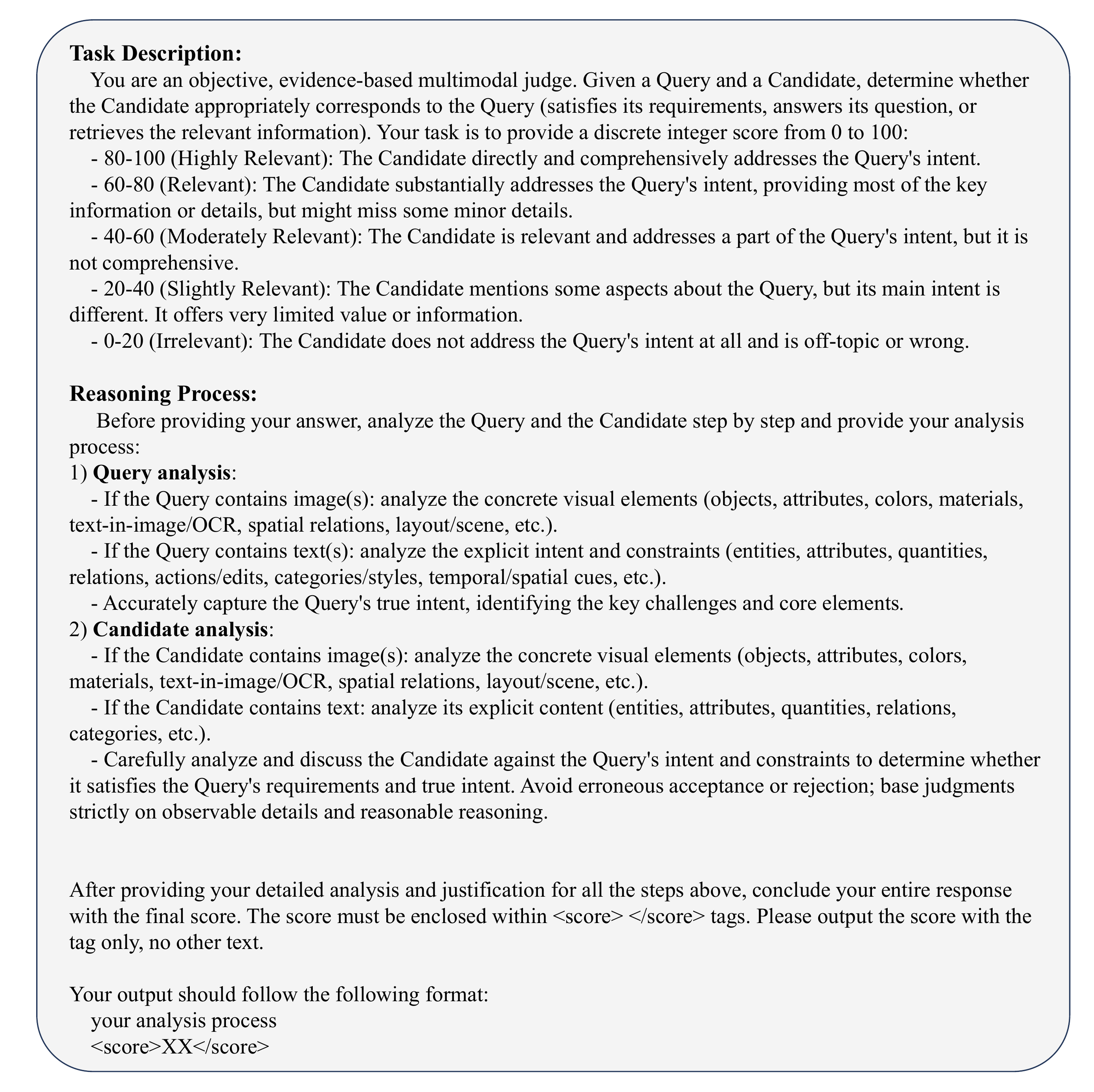}
    \caption{Prompt used by MLLMs to score query-candidate pairs after reasoning.}
    \label{fig:prompt_reranker}
\end{figure}

For text-only rerankers, following the second input configuration described in~\Cref{appendix:baselines}, we append image captions as auxiliary context. For multimodal rerankers, MLLMs are prompted in a \textit{reason-then-rank} format; the full prompt is provided in~\Cref{fig:prompt_reranker}. We evaluate GPT-5 via its official API~\footnote{gpt-5-2025-08-07}, and BGE-Reasoner-Reranker-32B with the authors’ code and checkpoint obtained via email. For open-source MLLMs (Gemma-3-27B, Qwen2.5-VL-72B, GLM-4.5V), we run inference with SGLang~\footnote{\href{[https://docs.sglang.ai/}{https://docs.sglang.ai/}} to accelerate the reasoning stage. All other models are evaluated using their released code and checkpoints.

\subsection{Detailed Results}
\label{appendix: rerank reuslts}
We report detailed reranking results for three retrievers (Qwen3-Embedding, GME, and Seed-1.6-Embedding) in ~\Cref{tab:detailed-reranker-results-qwen3}, ~\Cref{tab:detailed-reranker-results-gme}, and ~\Cref{tab:detailed-reranker-results-seed}, respectively.

\definecolor{caprow}{HTML}{F2F2F2}
\begin{table}[htbp]
\begin{adjustbox}{max width=\textwidth}
\renewcommand{\arraystretch}{1.10}
\setlength{\extrarowheight}{0.2em}
\begin{tabular}{@{}cccccccccccccc@{}}
\toprule
\multicolumn{1}{c|}{\multirow{2}{*}{\textbf{Methods}}} &
\multicolumn{6}{c|}{\textbf{Multimodal Knowledge Retrieval}} &
\multicolumn{3}{c|}{\textbf{Visual Illustration}} &
\multicolumn{3}{c|}{\textbf{Visual Relation}} &
\multicolumn{1}{l}{\multirow{2}{*}{\textbf{Avg.}}} \\
\cmidrule(lr){2-7} \cmidrule(lr){8-10} \cmidrule(lr){11-13}
\multicolumn{1}{c|}{} & \textbf{Bio.} & \textbf{Cook.} & \textbf{Gar.} & \textbf{Phy.} & \textbf{Chem.} & \multicolumn{1}{c|}{\textbf{Earth.}} &
\textbf{Econ.} & \textbf{Math.} & \multicolumn{1}{c|}{\textbf{Nat.}} &
\textbf{Spa.} & \textbf{Puzz.} & \multicolumn{1}{c|}{\textbf{Ana.}} &
\multicolumn{1}{l}{} \\
\midrule
\multicolumn{14}{c}{\textcolor{gray}{\textbf{\textit{Base Retriever}}}} \\ \midrule
\multicolumn{1}{l|}{Qwen3-Embedding} & {29.97} & {29.29} & {18.32} & {21.46} & {9.52} & \multicolumn{1}{c|}{{23.19}} & {49.44} & {21.14} & \multicolumn{1}{c|}{{26.30}} & {9.11} & {0.00} & \multicolumn{1}{c|}{{4.30}} & {20.17} \\
\midrule
\multicolumn{14}{c}{\textcolor{gray}{\textbf{\textit{Textual Rerankers}}}} \\ \midrule
\multicolumn{1}{l|}{{RankLLaMa-7B}} & 30.58 & 28.35 & 14.86 & 23.71 & 10.44 & \multicolumn{1}{c|}{26.29} & 48.66 & 13.48 & \multicolumn{1}{c|}{34.31} & 11.03 & 0.00 & \multicolumn{1}{c|}{9.39} & 20.92 \\
\multicolumn{1}{l|}{{RankLLaMa-14B}} & 33.92 & 32.62 & 13.82 & 22.35 & 11.98 & \multicolumn{1}{c|}{32.82} & 40.31 & 15.34 & \multicolumn{1}{c|}{29.22} & 15.76 & 0.00 & \multicolumn{1}{c|}{8.57} & 21.39 \\
\multicolumn{1}{l|}{{Rank1-7B}} & 32.05 & 37.15 & 17.50 & 20.11 & 12.04 & \multicolumn{1}{c|}{30.06} & 55.96 & 19.14 & \multicolumn{1}{c|}{39.81} & 16.41 & 0.00 & \multicolumn{1}{c|}{6.87} & 23.92 \\
\multicolumn{1}{l|}{{RankR1-14B}} & 35.39 & 35.87 & 19.49 & 23.89 & 12.38 & \multicolumn{1}{c|}{29.86} & 59.66 & 16.84 & \multicolumn{1}{c|}{40.43} & 20.26 & 0.00 & \multicolumn{1}{c|}{10.54} & 25.38 \\
\multicolumn{1}{l|}{{ReasonRank-32B}} & 34.49 & 36.50 & 20.02 & 23.49 & 12.45 & \multicolumn{1}{c|}{30.21} & 59.08 & 18.86 & \multicolumn{1}{c|}{37.60} & 17.25 & 0.00 & \multicolumn{1}{c|}{11.32} & 25.11 \\
\multicolumn{1}{l|}{{BGE-Reasoner-Reranker-32B}} & 37.05 & 40.29 & 21.98 & 22.33 & 14.24 & \multicolumn{1}{c|}{32.43} & 62.27 & 18.45 & \multicolumn{1}{c|}{44.00} & 24.13 & 0.00 & \multicolumn{1}{c|}{11.24} & 27.37 \\
\midrule
\multicolumn{14}{c}{\textcolor{gray}{\textbf{\textit{Multimodal Rerankers}}}} \\ \midrule
\multicolumn{1}{l|}{{MonoQwen2-VL}} & 31.61 & 35.58 & 17.13 & 19.82 & 9.49 & \multicolumn{1}{c|}{23.25} & 60.35 & 14.81 & \multicolumn{1}{c|}{55.25} & 13.87 & 0.24 & \multicolumn{1}{c|}{11.42} & 24.40 \\
\multicolumn{1}{l|}{{Jina-Reranker}} & 31.97 & 36.23 & 19.39 & 19.56 & 8.92 & \multicolumn{1}{c|}{23.78} & 63.58 & 17.20 & \multicolumn{1}{c|}{54.47} & 23.24 & 0.39 & \multicolumn{1}{c|}{10.29} & 25.84 \\
\multicolumn{1}{l|}{{Gemma-3-27B}} & 36.20 & 42.07 & 19.72 & 21.16 & 14.82 & \multicolumn{1}{c|}{27.93} & 49.19 & 19.01 & \multicolumn{1}{c|}{44.77} & 26.94 & 0.00 & \multicolumn{1}{c|}{16.21} & 26.50 \\
\multicolumn{1}{l|}{{Qwen2.5-VL-72B}} & 35.36 & 38.93 & 21.18 & 20.97 & 13.56 & \multicolumn{1}{c|}{30.49} & 57.23 & 21.41 & \multicolumn{1}{c|}{49.84} & 26.62 & 0.20 & \multicolumn{1}{c|}{16.37} & 27.68 \\
\multicolumn{1}{l|}{{GLM-4.5V}} & 35.60 & 41.26 & 18.95 & 21.10 & 14.53 & \multicolumn{1}{c|}{28.70} & 55.39 & 20.21 & \multicolumn{1}{c|}{52.55} & 30.73 & 0.62 & \multicolumn{1}{c|}{17.83} & 28.12 \\
\bottomrule
\end{tabular}
\end{adjustbox}
\caption{{Detailed reranking performance (nDCG@10) on MR$^{2}$-Bench with Qwen3-Embedding as the base retriever.}}
\label{tab:detailed-reranker-results-qwen3}
\end{table}

\definecolor{caprow}{HTML}{F2F2F2}
\begin{table}[htbp]
\begin{adjustbox}{max width=\textwidth}
\renewcommand{\arraystretch}{1.10}
\setlength{\extrarowheight}{0.2em}
\begin{tabular}{@{}cccccccccccccc@{}}
\toprule
\multicolumn{1}{c|}{\multirow{2}{*}{\textbf{Methods}}} &
\multicolumn{6}{c|}{\textbf{Multimodal Knowledge Retrieval}} &
\multicolumn{3}{c|}{\textbf{Visual Illustration}} &
\multicolumn{3}{c|}{\textbf{Visual Relation}} &
\multicolumn{1}{l}{\multirow{2}{*}{\textbf{Avg.}}} \\
\cmidrule(lr){2-7} \cmidrule(lr){8-10} \cmidrule(lr){11-13}
\multicolumn{1}{c|}{} & \textbf{Bio.} & \textbf{Cook.} & \textbf{Gar.} & \textbf{Phy.} & \textbf{Chem.} & \multicolumn{1}{c|}{\textbf{Earth.}} &
\textbf{Econ.} & \textbf{Math.} & \multicolumn{1}{c|}{\textbf{Nat.}} &
\textbf{Spa.} & \textbf{Puzz.} & \multicolumn{1}{c|}{\textbf{Ana.}} &
\multicolumn{1}{l}{} \\
\midrule
\multicolumn{14}{c}{\textcolor{gray}{\textbf{\textit{Base Retriever}}}} \\ \midrule
\multicolumn{1}{l|}{GME} & 34.34 & 39.50 & 19.04 & 19.29 & 7.73 & \multicolumn{1}{c|}{28.59} & 36.95 & 7.19 & \multicolumn{1}{c|}{39.35} & 15.70 & 0.22 & \multicolumn{1}{c|}{11.11} & 21.59 \\
\midrule
\multicolumn{14}{c}{\textcolor{gray}{\textbf{\textit{Textual Rerankers}}}} \\ \midrule
\multicolumn{1}{l|}{{RankLLaMa-7B}} & 30.58 & 28.35 & 14.86 & 23.71 & 10.44 & \multicolumn{1}{c|}{26.29} & 48.66 & 13.48 & \multicolumn{1}{c|}{34.31} & 11.03 & 0.00 & \multicolumn{1}{c|}{9.39} & 20.92 \\
\multicolumn{1}{l|}{{RankLLaMa-14B}} & 33.92 & 32.62 & 13.82 & 22.35 & 11.98 & \multicolumn{1}{c|}{32.82} & 40.31 & 15.34 & \multicolumn{1}{c|}{29.22} & 15.76 & 0.00 & \multicolumn{1}{c|}{8.57} & 21.39 \\
\multicolumn{1}{l|}{{Rank1-7B}} & 32.05 & 37.15 & 17.50 & 20.11 & 12.04 & \multicolumn{1}{c|}{30.06} & 55.96 & 19.14 & \multicolumn{1}{c|}{39.81} & 16.41 & 0.00 & \multicolumn{1}{c|}{6.87} & 23.92 \\
\multicolumn{1}{l|}{{RankR1-14B}} & 35.39 & 35.87 & 19.49 & 23.89 & 12.38 & \multicolumn{1}{c|}{29.86} & 59.66 & 16.84 & \multicolumn{1}{c|}{40.43} & 20.26 & 0.00 & \multicolumn{1}{c|}{10.54} & 25.38 \\
\multicolumn{1}{l|}{{ReasonRank-32B}} & 34.49 & 36.50 & 20.02 & 23.49 & 12.45 & \multicolumn{1}{c|}{30.21} & 59.08 & 18.86 & \multicolumn{1}{c|}{37.60} & 17.25 & 0.00 & \multicolumn{1}{c|}{11.32} & 25.11 \\
\multicolumn{1}{l|}{{BGE-Reasoner-Reranker-32B}} & 37.05 & 40.29 & 21.98 & 22.33 & 14.24 & \multicolumn{1}{c|}{32.43} & 62.27 & 18.45 & \multicolumn{1}{c|}{44.00} & 24.13 & 0.00 & \multicolumn{1}{c|}{11.24} & 27.37 \\
\midrule
\multicolumn{14}{c}{\textcolor{gray}{\textbf{\textit{Multimodal Rerankers}}}} \\ \midrule
\multicolumn{1}{l|}{{MonoQwen2-VL}} & 31.61 & 35.58 & 17.13 & 19.82 & 9.49 & \multicolumn{1}{c|}{23.25} & 60.35 & 14.81 & \multicolumn{1}{c|}{55.25} & 13.87 & 0.24 & \multicolumn{1}{c|}{11.42} & 24.40 \\
\multicolumn{1}{l|}{{Jina-Reranker}} & 31.97 & 36.23 & 19.39 & 19.56 & 8.92 & \multicolumn{1}{c|}{23.78} & 63.58 & 17.20 & \multicolumn{1}{c|}{54.47} & 23.24 & 0.39 & \multicolumn{1}{c|}{10.29} & 25.84 \\
\multicolumn{1}{l|}{{Gemma-3-27B}} & 36.20 & 42.07 & 19.72 & 21.16 & 14.82 & \multicolumn{1}{c|}{27.93} & 49.19 & 19.01 & \multicolumn{1}{c|}{44.77} & 26.94 & 0.00 & \multicolumn{1}{c|}{16.21} & 26.50 \\
\multicolumn{1}{l|}{{Qwen2.5-VL-72B}} & 35.36 & 38.93 & 21.18 & 20.97 & 13.56 & \multicolumn{1}{c|}{30.49} & 57.23 & 21.41 & \multicolumn{1}{c|}{49.84} & 26.62 & 0.20 & \multicolumn{1}{c|}{16.37} & 27.68 \\
\multicolumn{1}{l|}{{GLM-4.5V}} & 35.60 & 41.26 & 18.95 & 21.10 & 14.53 & \multicolumn{1}{c|}{28.70} & 55.39 & 20.21 & \multicolumn{1}{c|}{52.55} & 30.73 & 0.62 & \multicolumn{1}{c|}{17.83} & 28.12 \\
\bottomrule
\end{tabular}
\end{adjustbox}
\caption{{Detailed reranking performance (nDCG@10) on MR$^{2}$-Bench with GME as the base retriever.}}
\label{tab:detailed-reranker-results-gme}
\end{table}

\definecolor{caprow}{HTML}{F2F2F2}
\begin{table}[htbp]
\begin{adjustbox}{max width=\textwidth}
\renewcommand{\arraystretch}{1.10}
\setlength{\extrarowheight}{0.2em}
\begin{tabular}{@{}cccccccccccccc@{}}
\toprule
\multicolumn{1}{c|}{\multirow{2}{*}{\textbf{Methods}}} &
\multicolumn{6}{c|}{\textbf{Multimodal Knowledge Retrieval}} &
\multicolumn{3}{c|}{\textbf{Visual Illustration}} &
\multicolumn{3}{c|}{\textbf{Visual Relation}} &
\multicolumn{1}{l}{\multirow{2}{*}{\textbf{Avg.}}} \\
\cmidrule(lr){2-7} \cmidrule(lr){8-10} \cmidrule(lr){11-13}
\multicolumn{1}{c|}{} & \textbf{Bio.} & \textbf{Cook.} & \textbf{Gar.} & \textbf{Phy.} & \textbf{Chem.} & \multicolumn{1}{c|}{\textbf{Earth.}} &
\textbf{Econ.} & \textbf{Math.} & \multicolumn{1}{c|}{\textbf{Nat.}} &
\textbf{Spa.} & \textbf{Puzz.} & \multicolumn{1}{c|}{\textbf{Ana.}} &
\multicolumn{1}{l}{} \\
\midrule
\multicolumn{14}{c}{\textcolor{gray}{\textbf{\textit{Base Retriever}}}} \\ \midrule
\multicolumn{1}{l|}{Seed-1.6-Embedding} & 40.64 & 38.12 & 31.77 & 27.91 & 17.80 & \multicolumn{1}{c|}{37.17} & 56.13 & 26.10 & \multicolumn{1}{c|}{65.16} & 17.29 & 0.93 & \multicolumn{1}{c|}{9.21} & 30.68 \\
\midrule
\multicolumn{14}{c}{\textcolor{gray}{\textbf{\textit{Textual Rerankers}}}} \\ \midrule
\multicolumn{1}{l|}{{RankLLaMa-7B}} & 37.92 & 34.53 & 30.40 & 27.74 & 16.31 & \multicolumn{1}{c|}{30.35} & 54.62 & 32.71 & \multicolumn{1}{c|}{40.76} & 15.21 & 1.04 & \multicolumn{1}{c|}{6.42} & 27.33 \\
\multicolumn{1}{l|}{{RankLLaMa-14B}} & 43.27 & 38.94 & 29.61 & 26.92 & 20.32 & \multicolumn{1}{c|}{37.03} & 45.98 & 34.65 & \multicolumn{1}{c|}{35.08} & 16.62 & 1.49 & \multicolumn{1}{c|}{7.20} & 28.09 \\
\multicolumn{1}{l|}{{Rank1-7B}} & 36.95 & 35.07 & 24.49 & 25.34 & 21.21 & \multicolumn{1}{c|}{33.87} & 67.19 & 45.01 & \multicolumn{1}{c|}{47.22} & 17.39 & 1.96 & \multicolumn{1}{c|}{6.29} & 30.21 \\
\multicolumn{1}{l|}{{RankR1-14B}} & 40.11 & 34.96 & 26.88 & 28.58 & 19.60 & \multicolumn{1}{c|}{34.36} & 77.49 & 39.56 & \multicolumn{1}{c|}{46.30} & 24.41 & 1.93 & \multicolumn{1}{c|}{13.19} & 32.28 \\
\multicolumn{1}{l|}{{ReasonRank-32B}} & 41.70 & 37.31 & 28.98 & 28.34 & 18.61 & \multicolumn{1}{c|}{35.57} & 72.19 & 43.53 & \multicolumn{1}{c|}{45.35} & 21.10 & 2.30 & \multicolumn{1}{c|}{15.16} & 32.51 \\
\multicolumn{1}{l|}{{BGE-Reasoner-Reranker-32B}} & 45.19 & 39.31 & 32.18 & 28.57 & 20.69 & \multicolumn{1}{c|}{39.26} & 76.53 & 44.35 & \multicolumn{1}{c|}{53.07} & 28.35 & 2.09 & \multicolumn{1}{c|}{11.31} & 35.08 \\
\midrule
\multicolumn{14}{c}{\textcolor{gray}{\textbf{\textit{Multimodal Rerankers}}}} \\ \midrule
\multicolumn{1}{l|}{{MonoQwen2-VL}} & 35.33 & 36.41 & 24.71 & 23.96 & 15.31 & \multicolumn{1}{c|}{27.96} & 70.60 & 38.93 & \multicolumn{1}{c|}{64.83} & 18.23 & 1.14 & \multicolumn{1}{c|}{12.28} & 30.64 \\
\multicolumn{1}{l|}{{Jina-Reranker}} & 34.23 & 35.45 & 28.13 & 24.25 & 15.67 & \multicolumn{1}{c|}{25.86} & 79.48 & 43.21 & \multicolumn{1}{c|}{63.63} & 31.90 & 0.60 & \multicolumn{1}{c|}{11.35} & 32.82 \\
\multicolumn{1}{l|}{{Gemma-3-27B}} & 39.94 & 39.67 & 26.15 & 25.57 & 25.11 & \multicolumn{1}{c|}{33.94} & 59.75 & 44.20 & \multicolumn{1}{c|}{54.44} & 34.99 & 1.96 & \multicolumn{1}{c|}{16.44} & 33.51 \\
\multicolumn{1}{l|}{{Qwen2.5-VL-72B}} & 42.95 & 38.78 & 29.60 & 28.21 & 21.17 & \multicolumn{1}{c|}{37.66} & 72.57 & 51.09 & \multicolumn{1}{c|}{61.47} & 31.97 & 4.58 & \multicolumn{1}{c|}{14.29} & 36.20 \\
\multicolumn{1}{l|}{{GLM-4.5V-thinking}} & 42.43 & 41.28 & 26.37 & 29.78 & 24.34 & \multicolumn{1}{c|}{34.15} & 70.52 & 50.24 & \multicolumn{1}{c|}{60.24} & 36.06 & 3.78 & \multicolumn{1}{c|}{17.29} & 36.37 \\
\multicolumn{1}{l|}{{GPT-5}} & 52.35 & 55.41 & 37.46 & 37.10 & 31.96 & \multicolumn{1}{c|}{51.12} & 83.83 & 55.63 & \multicolumn{1}{c|}{79.16} & 41.41 & 3.94 & \multicolumn{1}{c|}{21.48} & 45.90 \\
\bottomrule
\end{tabular}
\end{adjustbox}
\caption{{Detailed reranking performance (nDCG@10) on MR$^{2}$-Bench with Seed-1.6-Embedding as the base retriever.}}
\label{tab:detailed-reranker-results-seed}
\end{table}

\end{document}